\documentclass[%
 superscriptaddress,
 showkeys,
]{revtex4-2}

\usepackage{chngcntr}

\usepackage{hyperref}       
\usepackage{url}            
\usepackage{booktabs}       
\usepackage{amsfonts}       
\usepackage{nicefrac}       
\usepackage{microtype}      
\usepackage{algpseudocode}
\usepackage{textcomp}
\usepackage{amssymb}
\usepackage{amsmath}
\usepackage{bm}
\usepackage{caption, subcaption}
\usepackage{anyfontsize}
\usepackage{etoolbox}
\usepackage{cleveref}
\usepackage{wasysym}
\usepackage[export]{adjustbox}
\usepackage{graphicx}
\usepackage{mathtools}
\usepackage[normalem]{ulem}
\usepackage{xcolor}
\usepackage{soul}

\DeclarePairedDelimiter{\ceil}{\lceil}{\rceil}

\newcommand{\decoder}{
\mathcal{D}^{\bm{w}_{ \mathcal{D}}}
}

\newcommand{\encoder}{
\mathcal{E}^{\bm{w}_{ \mathcal{E}}}
}

\newcommand{\calL}{\mathcal{L}}
\newcommand{\calR}{\mathcal{R}}
\newcommand{\calH}{\mathcal{H}}

\newcommand{\calD}{\mathcal{D}}
\newcommand{\calN}{\mathcal{N}}

\newcommand{\bmtx}{\bm{\tilde{x}}}

\newcommand{\bmtz}{\bm{\tilde{z}}}
\newcommand{\bmth}{\bm{\tilde{h}}}

\newcommand{\bmw}{\bm{w}}
\newcommand{\bmv}{\bm{v}}
\newcommand{\bmu}{\bm{u}}
\newcommand{\bmx}{\bm{x}}
\newcommand{\bms}{\bm{s}}
\newcommand{\bmz}{\bm{z}}

\newcommand{\bmh}{\bm{h}}

\newcommand{\bmr}{\bm{r}}
\newcommand{\bmb}{\bm{b}}
\newcommand{\bmg}{\bm{g}}
\newcommand{\bmc}{\bm{c}}

\definecolor{darkgreen}{rgb}{0.0, 0.5, 0.0}

\graphicspath{{./Figures/}}

\newcommand{\addlegend}[2]{%
{\raisebox{-0.3ex}{\includegraphics[scale=#2]{./Legend/#1}}}}

\begin{document}
\title{
Multiscale Simulations of Complex Systems by Learning their Effective Dynamics
}

\author{Pantelis R.~Vlachas}
\affiliation{Computational Science and Engineering Laboratory, ETH Z\"urich, CH-8092, Switzerland}
\affiliation{School of Engineering and Applied Sciences, 29 Oxford Street, Harvard University, Cambridge, MA 02138, USA}

\author{Georgios Arampatzis}
\affiliation{Computational Science and Engineering Laboratory, ETH Z\"urich, CH-8092, Switzerland}
\affiliation{School of Engineering and Applied Sciences, 29 Oxford Street, Harvard University, Cambridge, MA 02138, USA}

\author{Caroline Uhler}
\affiliation{Institute for Data, Systems, and Society, Massachusetts Institute of Technology, 77 Massachusetts Avenue, Cambridge, MA 02139, USA}

\author{Petros Koumoutsakos}\thanks{Corresponding author, petros@seas.harvard.edu}
\affiliation{Computational Science and Engineering Laboratory, ETH Z\"urich, CH-8092, Switzerland}
\affiliation{School of Engineering and Applied Sciences, 29 Oxford Street, Harvard University, Cambridge, MA 02138, USA}

\date{\today}

\begin{abstract}
Predictive simulations of complex systems are essential for applications ranging from weather forecasting to drug design.
The veracity of these predictions hinges on their capacity to capture the effective system dynamics. 
Massively parallel simulations predict the system dynamics by resolving all spatiotemporal scales, often at a cost that prevents experimentation while their findings may not allow for generalisation.
On the other hand reduced order models are fast but limited by the frequently adopted linearization of the system dynamics and/or the utilization of heuristic closures.
Here we present a novel systematic framework that bridges large scale simulations and reduced order models to Learn the Effective Dynamics (LED) of diverse complex systems. 
The framework forms algorithmic alloys between non-linear machine learning algorithms and the  Equation-Free approach for modeling complex systems.
LED deploys autoencoders to formulate a mapping between fine and coarse-grained representations and evolves the latent space dynamics using recurrent neural networks.
The algorithm is validated on benchmark problems and we find that it outperforms state of the art reduced order models in terms of predictability and large scale simulations in terms of cost.
LED is applicable to systems ranging from chemistry to fluid mechanics and reduces the computational effort by up to two orders of magnitude while maintaining the prediction accuracy of the full system dynamics.
We argue that LED provides a novel potent modality for the accurate prediction of complex systems.
\end{abstract}


\maketitle

Some of the most important scientific advances and engineering designs are founded on the study of complex systems that exhibit dynamics spanning multiple spatiotemporal scales.
Examples include protein dynamics~\cite{Rackovsky2020}, morphogenesis~\cite{Gilmour2017}, brain dynamics~\cite{Robinson2005}, climate~\cite{climatenas}, ocean dynamics~\cite{mahadevan2016impact} and social systems~\cite{bellomo2011modeling}.
Over the last fifty years, simulations have become a key component of these studies thanks to a confluence of advances in computing architectures, numerical methods and software.
Large scale simulations have led to unprecedented insight, acting as in-silico microscopes~\cite{lee2009discovery} or telescopes to reveal the dynamics of galaxy formations~\cite{springel2005simulations}.
At the same time these simulations have led to the understanding that resolving the full range of spatio-temporal scales in such complex systems will remain out of reach in the foreseeable future.

In recent years there have been intense efforts to develop efficient simulations that exploit the multiscale character of the systems under investigation~\cite{car1985unified,kevrekidis2003equation, weinan2003heterognous,kevrekidis2004equation}.
Multiscale methods rely on judicious approximations of the interactions between processes occurring over different scales and a number of potent frameworks have been proposed including the Equation-Free framework (EFF) ~\cite{kevrekidis2003equation,kevrekidis2004equation,laing2010reduced,bar2019learning}, the Heterogeneous Multiscale Method (HMM)~\cite{weinan2003heterognous,Weinan2004,weinan2007heterogeneous}, and the FLow AVeraged integatoR (FLAVOR)~\cite{tao2010nonintrusive}.
In these algorithms the system dynamics are distinguished into fine and coarse scales or expensive and affordable simulations, respectively.
Their success depends on the separation of scales that are inherent to the system dynamics and their capability to capture the transfer of information between scales.
Effective applications of multiscale methodologies minimize the computational effort while maximizing the accuracy of the propagated dynamics.
The EFF relies on few fine scale simulations that are used to acquire, through ``restricting'', information about the evolution of the coarse-grained quantities of interest.
In turn various time stepping procedures are used to propagate the coarse-grained dynamics.
The fine scale dynamics are obtained by judiciously ``lifting'' the coarse scales to return to the fine scale description of the system and repeat.
When the EFF reproduces trajectories of the original system, the identified low order dynamics represent the intrinsic system dynamics, also called effective dynamics, inertial manifold~\cite{linot2020deep,robinson1994inertial} or reaction coordinates in molecular kinetics.

While it is undisputed that the EFF, HMM, FLAVOR and related frameworks have revolutionized the field of multiscale modeling and simulation, we identify two critical issues that presently limit their potential.
First, the accuracy of propagating the coarse-grained/latent dynamics hinges on the employed time integrators.
Second, the choice of information transfer, in particular from coarse to fine scale dynamics in `lifting', greatly affects the forecasting capacity of the methods.

In the present work these two critical issues are resolved through machine learning (ML) algorithms that (i) deploy recurrent neural networks (RNNs) with gating mechanisms to evolve the coarse-grained dynamics and (ii) employ advanced (convolutional, or probabilistic) autoencoders (AE) to transfer in a systematic, data driven manner, the information between coarse and fine scale descriptions.

Over the last years, ML algorithms have exploited the ample availability of data, and powerful computing architectures, to provide us with remarkable successes across scientific disciplines~\cite{jumper2021highly, brunton2019machine}.
The particular elements of our algorithms have been employed in the modeling of dynamical systems.
Autoencoders (AEs) have been used to identify a linear latent space based on the Koopman framework~\cite{lusch2018deep}, model high-dimensional fluid flows~\cite{geneva2020modeling,milano2002neural} or sample effectively the state space in the kinetics of proteins~\cite{wehmeyer2018time}.
More recently AEs have been coupled with dynamic importance sampling~\cite{bhatia2021machine} to accelerate multiscale simulations and investigate the interactions of RAS proteins with a plasma membrane.
RNNs with gating mechanisms have been shown successful in a wide range of applications, from speech processing~\cite{chung2015recurrent} to complex systems~\cite{vlachas2020backpropagation}, but their effectiveness in a multiscale setting has yet to be investigated.
AEs coupled with RNNs are used in~\cite{gonzalez2018deep, maulik2021reduced, hasegawa2020machine} to model fluid flows.
In~\cite{lee2020coarse}, the authors build on the EFF framework, identify a PDE on a coarse representation by diffusion maps, Gaussian processes or neural networks, and utilize forward integration in the coarse representation.
These previous works, however, fail to employ one or more of the following mechanisms in contrast to our framework: consider the coarse scale dynamics~\cite{geneva2020modeling, milano2002neural}, account their non-Markovian~\cite{bhatia2021machine, lee2020coarse} or non-linear nature~\cite{lusch2018deep}, exploit a probabilistic generative mapping~\cite{geneva2020modeling, gonzalez2018deep, maulik2021reduced, hasegawa2020machine} from the coarse to the fine scale, learn simultaneously the latent space and its dynamics in an end-to-end fashion and not sequentially~\cite{geneva2020modeling, lee2020coarse, lusch2018deep, bhatia2021machine, gonzalez2018deep, maulik2021reduced, hasegawa2020machine}, alternate between micro and macro dynamics~\cite{geneva2020modeling, lusch2018deep, lee2020coarse, lusch2018deep, gonzalez2018deep, maulik2021reduced, hasegawa2020machine}, and scale to high-dimensional systems~\cite{lee2020coarse, gonzalez2018deep, maulik2021reduced}.

Augmenting multiscale frameworks (including EFF, HMM, FLAVOR) with state of the art ML algorithms allows for evolving the coarse scale dynamics by taking into account their time history and by providing consistent lifting (decoding) and restriction (encoding) operators to transfer information between fine and coarse scales.
We demonstrate that the proposed framework allows for simulations of complex multiscale systems that reduce the computational cost by orders of magnitude, to capture spatiotemporal scales that would be impossible to resolve with existing computing resources.

\section{Learning the Effective Dynamics (LED)}

We propose a framework for learning the effective dynamics (LED) of complex systems, that allows for accurate prediction of the system evolution at a significant reduced computational cost.

In the following, the high-dimensional state of a dynamical system is given by $\bm{s}_t \in \mathbb{R}^{d_{\bm{s}}}$, and the discrete time dynamics are given by
\begin{equation*}
\bm{s}_{t+\Delta t} = \bm{F}(\bm{s}_t),
\end{equation*}
where $\Delta t$ is the sampling period and $\bm{F}$ may be non-linear, deterministic or stochastic.
We assume that the 
state of the system at time $t$ can be described by a vector $\bm{z}_t \in \mathcal{Z}$, where $\mathcal{Z}\subset \mathbb{R}^{d_{\bm{z}}}$ is a low dimension manifold with $d_{\bm{z}} \ll d_{\bm{s}}$.
In order to identify this manifold, an encoder $\mathcal{E}^{\bm{w}_{ \mathcal{E}}}:\mathbb{R}^{d_{\bm{s}}} \to \mathbb{R}^{d_{\bm{z}}}$ is utilized, where $\bm{w}_{\mathcal{E}}$ are trainable parameters, transforming the high-dimensional state $\bm{s}_t$ to $\bm{z}_t = \mathcal{E}^{\bm{w}_{ \mathcal{E}}} (\bm{s}_t)$.
In turn, a decoder maps back this latent representation to the high-dimensional state, i.e. $\bm{\tilde{s}}_t=\mathcal{D}^{\bm{w}_{ \mathcal{D}}}(\bm{z}_t)$.

For deterministic systems, the optimal parameters  $\{\bm{w}_{\mathcal{E}}^{\star}, \bm{w}_{\mathcal{D}}^{\star} \}$ are identified by minimizing the mean squared reconstruction error (MSE),
\begin{gather*}
\bm{w}_{\mathcal{E}}^{\star}, \bm{w}_{\mathcal{D}}^{\star} =
\underset{\bm{w}_{\mathcal{E}}, \bm{w}_{\mathcal{D}}} {\operatorname{argmin}}
\Big (
\bm{s}_t - 
\bm{\tilde{s}}_t
\Big ) ^ 2
=
 \underset{\bm{w}_{\mathcal{E}}, \bm{w}_{\mathcal{D}}} {\operatorname{argmin}}
\Big (
\bm{s}_t - 
\mathcal{D}^{\bm{w}_{ \mathcal{D}}} \big(
\mathcal{E}^{\bm{w}_{ \mathcal{E}}}(\bm{s}_t)
\big )
\Big ) ^ 2
.
\end{gather*}
Convolutional neural network~\cite{lecun2015deep} autoencoders (CNN-AE) that take advantage of the spatial structure of the data are embedded into LED.

For stochastic systems, $\mathcal{D}^{\bm{w}_{ \mathcal{D}}}$ is modeled with a Mixture Density (MD) decoder~\cite{bishop1994mixture}.
Further details on the implementation of the MD decoder are provided in the SI, Section 1E, along with other components embedded in LED like AEs in SI, Section 1A, Variational AEs in SI, Section 1B, CNNs in SI, Section 1C.

We demonstrate the modularity of LED, as it can be coupled with a permutation invariant layer (see details in the SI, Section 1D), and utilized later in the modeling of the dynamics of a large set of particles governed by the advection diffusion equation (see details in the SI, Section 3A).

As a non-linear propagator in the low order manifold (coarse scale), an RNN is employed, capturing non-Markovian, memory effects by keeping an internal memory state.

The RNN is learning a forecasting rule
\begin{equation*}
\bm{h}_{t} =
\mathcal{H}^{
\bm{w}_{\mathcal{H}}}
\big(
\bm{z}_t, \bm{h}_{t-\Delta t}
\big),
\quad
\bm{\tilde{z}}_{t+\Delta t} =
\mathcal{R}^{
\bm{w}_{\mathcal{R}}
}
\big(
\bm{h}_t
\big),
\end{equation*}
where $\bm{h}_t \in \mathbb{R}^{d_h}$ is an internal hidden memory state, $\bm{\tilde{z}}_{t+\Delta t} $ is a latent state prediction,
$\mathcal{H}^{
\bm{w}_{\mathcal{H}}}$
and
$\mathcal{R}^{
\bm{w}_{\mathcal{R}}
}$
are the hidden-to-hidden, and the hidden-to-output mappings, and $\bm{w}_{\mathcal{H}}$, $\bm{w}_{\mathcal{R}}$ are the trainable parameters of the RNN.
One possible implementation of $\mathcal{H}^{
\bm{w}_{\mathcal{H}}}$
and
$\mathcal{R}^{
\bm{w}_{\mathcal{R}}}$
is the long short-term memory (LSTM)~\cite{Hochreiter1997}, presented in the SI, Section 1F.

\begin{figure}[tbhp]
\centering
\includegraphics[width=0.9\textwidth]{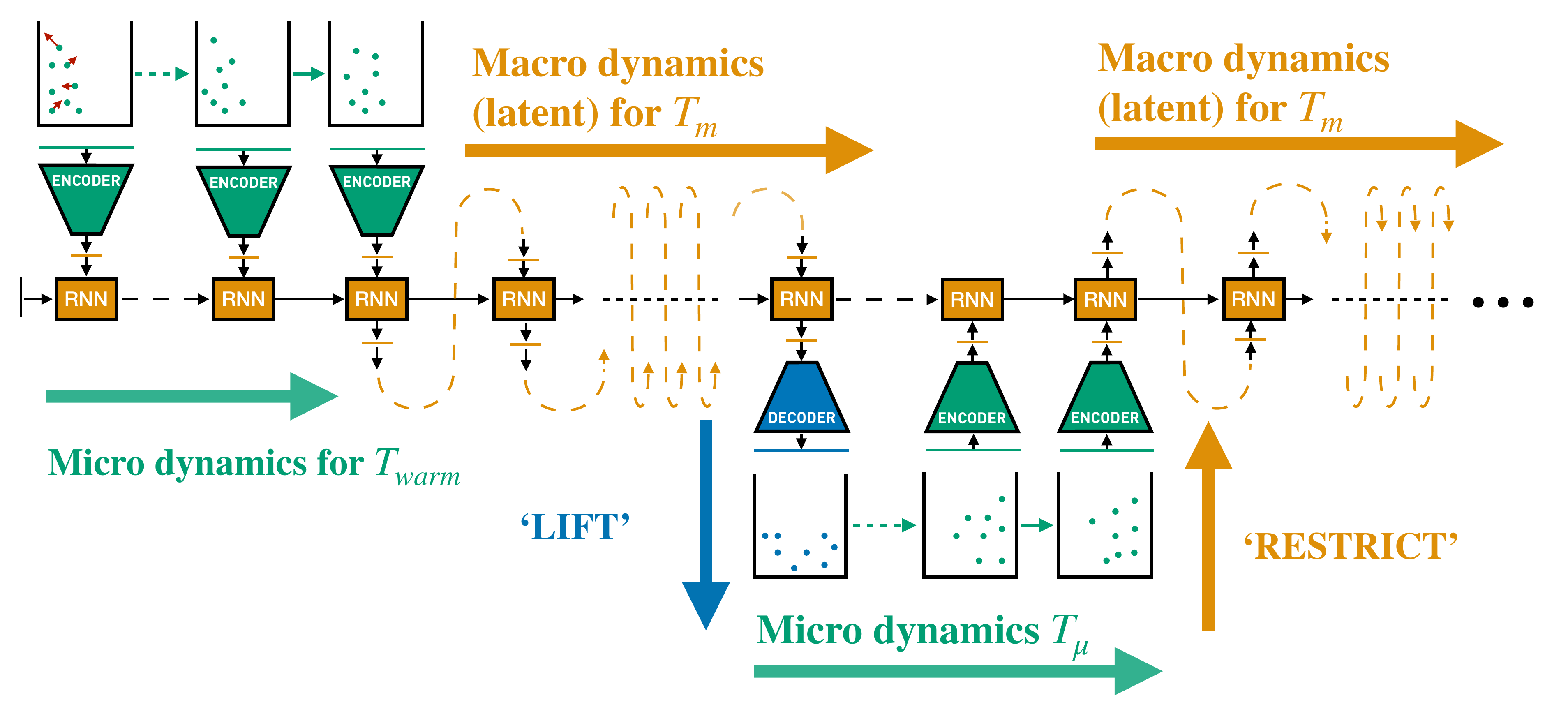}
\caption{
Multiscale-LED:
Starting from an initial condition use the equations/first principles to evolve the high-dimensional dynamics for a short period $T_{warm}$.
During this warm-up period, the state $\bm{s}_t$ is passed through the encoder network.
The outputs of the autoencoder are iteratively provided as inputs to the RNN, to warm-up its hidden state.
Next, iteratively, \textbf{(1)} starting from the last latent state $\bm{z}_t$ the RNN propagates the latent dynamics for $T_m \gg T_{warm}$, \textbf{(2)} lift the latent dynamics at $t=T_{warm} + T_{m}$ back to the high-dimensional state, \textbf{(3)} starting from this high-dimensional state as an initial condition, use the equations/first principles to evolve the dynamics for $T_{\mu} \ll T_m$.
}
\label{fig:LEDmultiscaleprop}
\end{figure}

The role of the RNN is twofold.
First, it is updating its hidden memory state $\bm{h}_t$, given the current state provided at the input $ \bm{z}_t$ and the hidden memory state at the previous time-step $\bm{h}_{t-\Delta t}$, tracking the history of the low order state to model non-Markovian dynamics.
Second, given the updated hidden state $\bm{h}_t$ the RNN forecasts the latent state at the next time-step(s) $\bm{\tilde{z}}_{t+\Delta t}$.
The RNN is trained to minimize the forecasting loss $||\bm{\tilde{z}}_{t+\Delta t} - \bm{z}_{t+\Delta t}||_2^2$ by backpropagation through time~\cite{Werbos1988}.

The LSTM and the AE, jointly referred to as LED, are trained on data from simulations of the fully resolved (or microscale) dynamical system.
The two networks can either be trained sequentially, or together.
In the first case, the AE is pretrained to minimize the reconstruction loss, and then the LSTM is trained to minimize the prediction loss on the latent space (AE-LSTM).
In the second case, they are seen as one network trying to minimize the sum of reconstruction and prediction losses (AE-LSTM-end2end).
For large, high-dimensional systems, the later approach of end-to-end training is computationally expensive.
After training, LED is employed to forecast the dynamics on unseen data, by propagating the low order latent state with the RNN and avoiding the computationally expensive simulation of high-dimensional dynamics.
We refer to this mode of propagation, iteratively propagating only the latent/macro dynamics, as Latent-LED.
We note that, as non-Markovian models are not self-starting, an initial small warm-up period is required, feeding the LED with data from the micro dynamics.

The LED framework allows for data driven information transfer between coarse and fine scales through the AE.
Moreover it propagates the latent space dynamics without the need to upscale back to the high-dimensional state space at every time-step.
As is the case for any approximate iterative integrator (here the RNN), the initial model errors will propagate.
In order to mitigate potential instabilities, inspired by the Equation-Free~\cite{kevrekidis2003equation}, we propose the multiscale forecasting scheme in~\Cref{fig:LEDmultiscaleprop}, alternating between micro dynamics for $T_{\mu}$ and macro dynamics for $T_m$.
In this way, the approximation error can be reduced at the cost of the computational complexity associated with evolving the high-dimensional dynamics.
We refer to this mode of propagation as Multiscale-LED, and the ratio $\rho=T_{m}/T_{\mu}$ as multiscale ratio.
In Multiscale-LED, the interface with the high-dimensional state space is enabled only at the time-steps and scales of interest.
This is in contrast to \cite{hernandez2018variational, sultan2018transferable}, and is easily adaptable to the needs of particular applications thus augmenting the arsenal of models developed for multiscale problems.

Training of LED models is performed with the Adam stochastic optimization method~\cite{Kingma2015}, and validation based early stopping is employed to avoid overfitting.
All LED models are implemented in Pytorch, mapped to a single Nvidia Tesla P100 GPU and executed on the XC50 compute nodes of the Piz Daint supercomputer at the Swiss national supercomputing centre (CSCS).

\section{Results}

We demonstrate the application of LED in a number of benchmark problems and compare its performance with existing state of the art algorithms.
In the SI Section 3D, we provide additional results on LED applied on alanine dipeptide in water.
The stochastic dynamics of the molecular system are handled with an MD decoder, and an MD-LSTM in the latent space~\cite{vlachas2021accelerated}.

\subsection{FitzHugh-Nagumo Model (FHN)}
\label{sec:fhnmodel}

LED is employed to capture the dynamics of the FitzHugh-Nagumo equations (FHN)~\cite{fitzhugh1961impulses,nagumo1962active}.
The FHN model describes the evolution of an activator $u(x,t)=\rho^{ac}(x,t)$ and an inhibitor density $v(x,t)=\rho^{in}(x,t)$ on the domain $x \in [ 0,L ]$:
\begin{equation}
\begin{aligned}
\frac{\partial u }{\partial t }
&=
D^{u} \frac{\partial^2 u }{\partial x^2}
+ u
- u^3
- v,
\\
\frac{\partial v }{\partial t }
&=
D^{v} \frac{\partial^2 v }{\partial x^2}
+ \epsilon
(
u - \alpha_1 v - \alpha_0
).
\label{eq:fhn}
\end{aligned}
\end{equation}
The system evolves periodically under two timescales, with the activator/inhibitor density acting as the ``fast''/``slow'' variable respectively.
The bifurcation parameter $\epsilon=0.006$ controls the difference in the time-scales.
We choose $D^{u}=1$, $D^{v}=4$, $L=20$, $\alpha_0=-0.03$ and $\alpha_1=2$.

\begin{figure}[tbhp]
\centering
\includegraphics[width=0.9\textwidth]{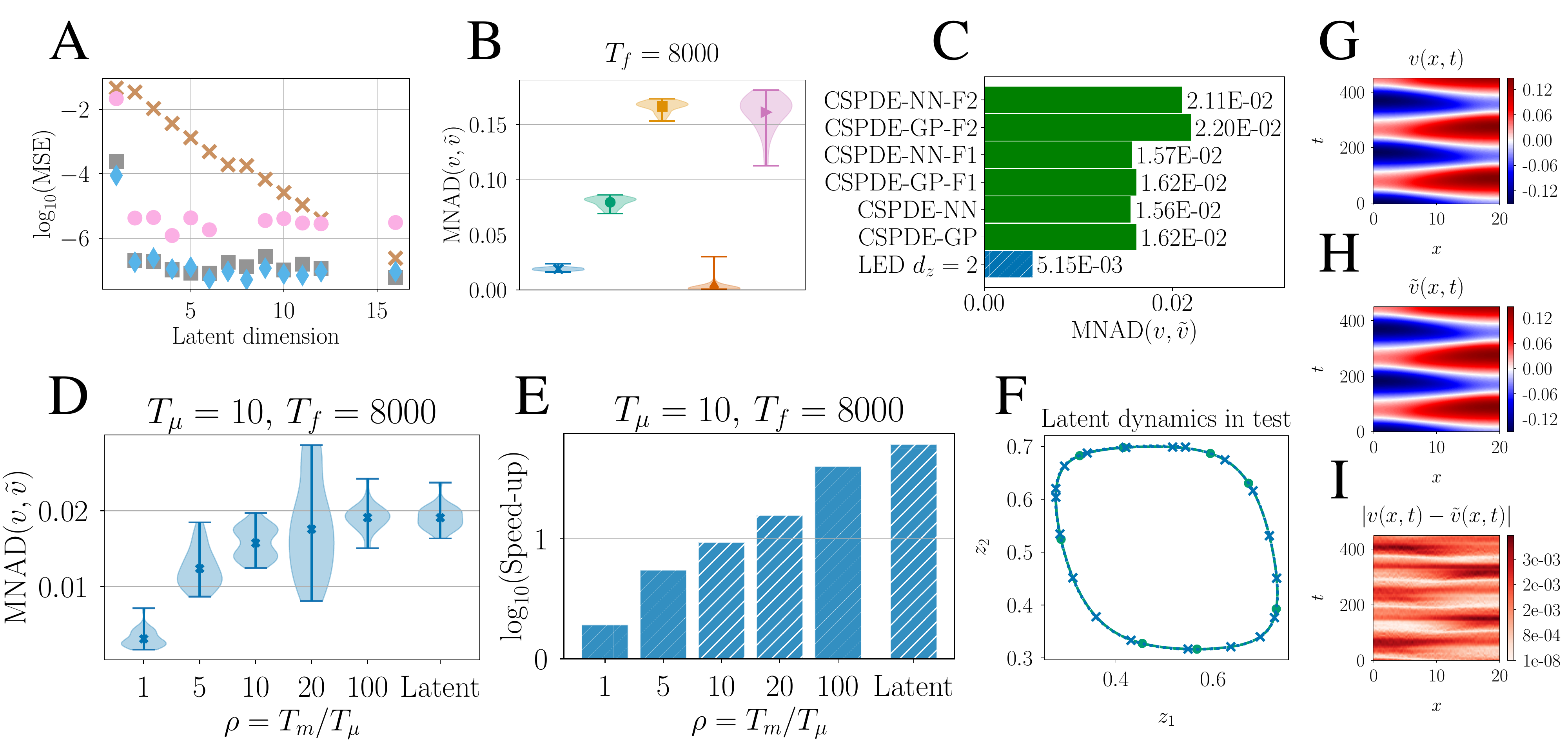}
\caption{
\textbf{A)} Comparison of the reconstruction MSE in the test data in the FHN between PCA (\addlegend{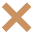}{0.5}), Diffusion Maps (\addlegend{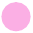}{0.5}), Autoencoder (\addlegend{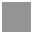}{0.5}), and Convolutional Autoencoder (\addlegend{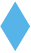}{0.5}).
\textbf{B)} Comparison of macrodynamics propagators (\addlegend{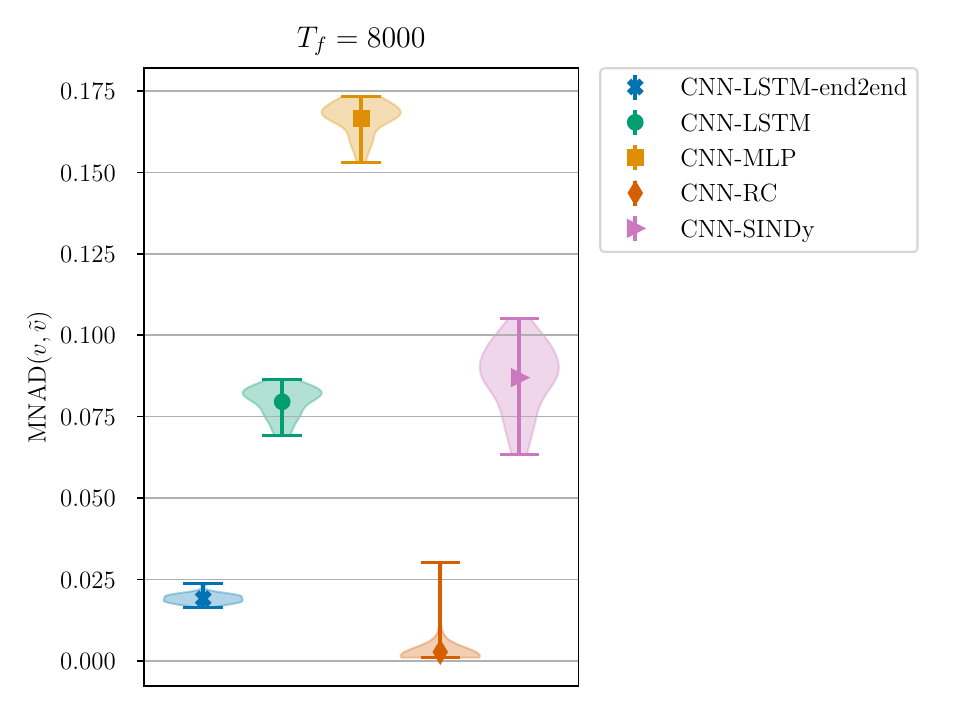}{0.95} AE-LSTM-end2end; \addlegend{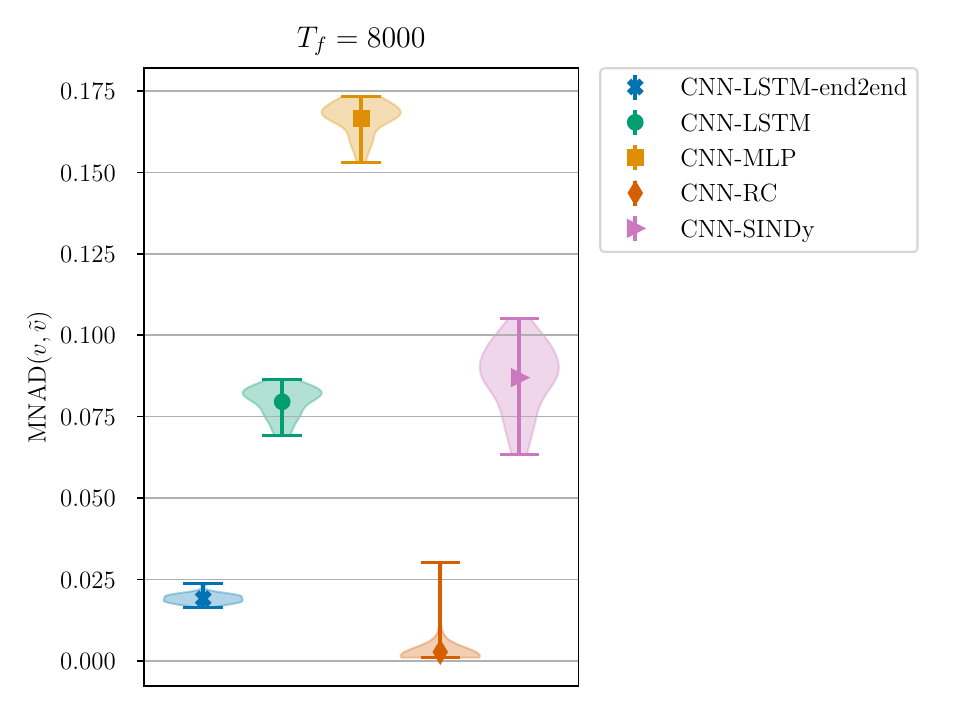}{0.95} AE-LSTM; \addlegend{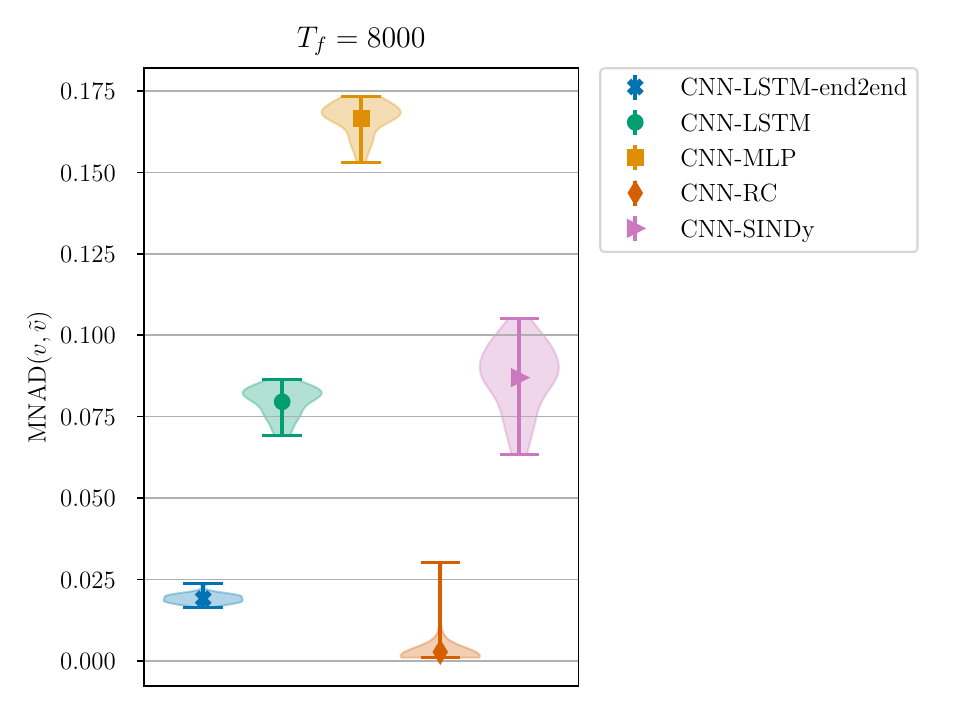}{0.95} AE-MLP; \addlegend{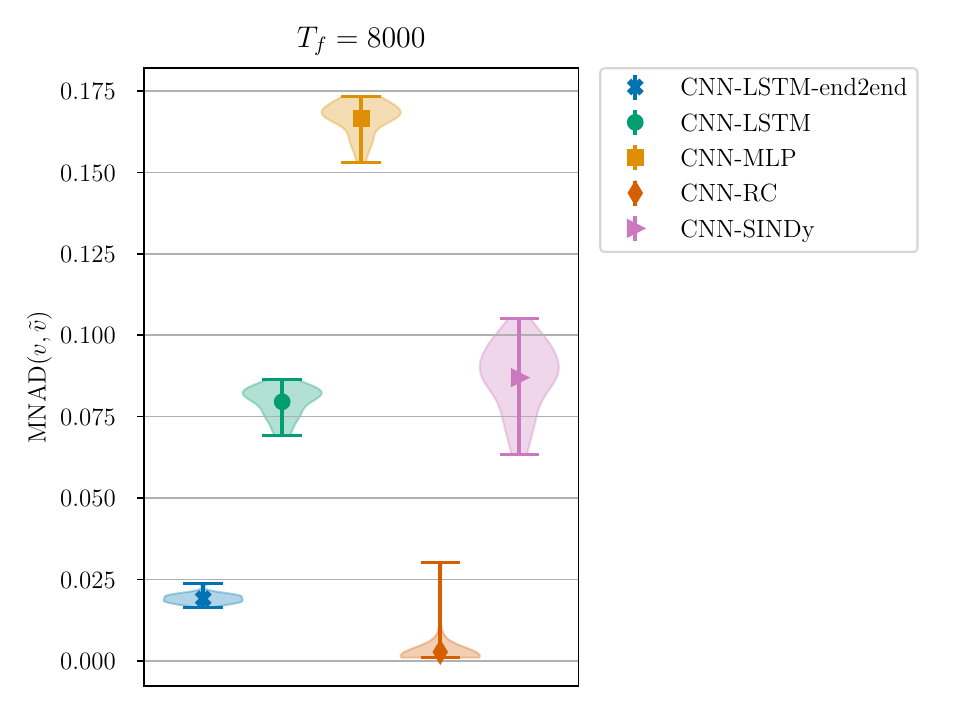}{0.95} AE-RC; \addlegend{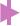}{0.95} AE-SINDy) in iterative latent forecasting.
\textbf{C)} Comparison of Latent-LED with $d_{\bm{z}}=2$ with Equation-Free variants from~\cite{lee2020coarse}.
\textbf{D)} The inhibitor MNAD of Multiscale-LED (AE-LSTM-end2end, $d_{\bm{z}}=2$) plotted as a function of the multiscale ratio $\rho=T_m/T_{\mu}$.
\textbf{E)} The speed-up of Multiscale-LED compared to the LB solver plotted w.r.t. $\rho$.
\textbf{F)} The LED latent state (\addlegend{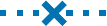}{0.72}) compared against the attractor of the data embedded in the latent space (\addlegend{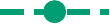}{0.72}).
\textbf{G)} A trajectory of the inhibitor density starting from a testing initial condition, along with \textbf{H)} the Latent-LED prediction,
and \textbf{I)} absolute difference.
}
\label{fig:FHN:FHN}
\end{figure}

\Cref{eq:fhn} is discretized with $N=101$ grid points and solved using the Lattice Boltzmann (LB) method~\cite{karlin2006elements}, with time-step $\delta_t =0.005$.
To facilitate comparison with~\cite{lee2020coarse}, we employ the LB method to gather data starting from $6$ different initial conditions to obtain the mesoscopic solution considered here as the fine-grained solution.
The data is sub-sampled, keeping every 200\textsuperscript{th} data point, i.e. the coarse time step is $\Delta t=1$.
Three time series with $451$ points are considered for training, two time series with $451$ points for validation, and $10^4$ data points from a different initial condition for testing.
For the identification of the latent space, we compare principal component analysis (PCA), diffusion maps, feed-forward AE, and CNN-AE, in terms of the mean squared error (MSE) of the reconstruction in the test data, plotted in~\Cref{fig:FHN:FHN} A.
The MSE is plateauing after $d_{\bm{z}}=2$, and the AE and CNN-AE exhibit at least an order of magnitude lower MSR compared to PCA and DiffMaps.
For this reason, we employ an AE with $d_{\bm{z}}=2$ for the LED.
The hyper-parameters of the networks (reported in the SI, Table 3 along with training times) are tuned based on the MSE on the validation data.
The architecture of the CNN is reported in Table 5, and depicted in Figure 10 of the SI.

In Figure \ref{fig:FHN:FHN} B, we compare various propagators in forecasting of the macro (latent dynamics), starting from $32$ different initial conditions in the test data, up to a horizon of $T_f=8000$.
We benchmark an AE-LSTM trained end-to-end (AE-LSTM-end2end), an AE-LSTM where the AE is pretrained (AE-LSTM), a multi-layered perceptron (AE-MLP), Reservoir Computers (AE-RC)~\cite{pathak2018model, vlachas2020backpropagation}, and the SINDy algorithm (AE-SINDy)~\cite{brunton2016discovering}.
As a comparison metric, we consider the mean normalised absolute difference (MNAD), averaged over the 32 initial conditions.
The definition of the MNAD is provided in SI, Section 2.
The MNAD is computed on the inhibitor density, as the difference between the result of the LB simulation $v(x,t)$, considered as groundtruth, and the model forecasts $\hat{v}$.
The warm-up period for all propagators is set to $T_{warm}=60$.
The hyper-parameters of the networks (reported in Tables 4, 6, and 7 of the SI, along with the training times) are tuned based on the MNAD on the validation data.
The LSTM-end2end and the RC show the lowest test error, while the variance of the RC is larger.
In the following, we consider an LSTM-end2end propagator for the LED.

LED is benchmarked against EFF variants ~\cite{lee2020coarse} in the FHN equation in Figure~\ref{fig:FHN:FHN} C.
As a metric for the accuracy, the MNAD is considered, consistent with~\cite{lee2020coarse} to facilitate comparison.
The EFF variants~\cite{lee2020coarse} are based on the identification of PDEs on the coarse level (CSPDE).
LED is compared with CSPDEs in forecasting the dynamics of the FHN equation starting from an initial condition from the test data up to final time $T_f=451$.
CSPDE variants are utilizing Gaussian processes (GP) or neural networks (NN), features of the fine scale dynamics obtained through diffusion maps (F1 to F3) and forward integration to propagate the coarse representation in time.
LED outperforms CSPDE variants by an order of magnitude.
In Figure~\ref{fig:FHN:FHN} F, the latent space of LED is plotted against the attractor of the data embedded in the latent space.
Even for long time horizons (here $T_f=8000$), the LED forecasts stay on the periodic attractor.

Latent-LED propagates the low order dynamics, and up-scales back to the inhibitor density, forecasting its evolution accurately, while being 60 times faster than the LB solver.
This speed-up can be decisive in accelerating simulations and achieving much larger time horizons.

In Multiscale-LED, the approximation error of LED decreases, at the cost of reduced speed-up.
This interplay can be seen in Figure~\ref{fig:FHN:FHN} D and E.

Latent-LED ($T_{\mu}=0$), and Multiscale-LED, alternating between macro-dynamics for $T_m=10$ and high-dimensional dynamics for $T_{\mu}$, are employed to approximate the evolution and compare it against the LB solver in forecasting up to $T_f=8000$ starting from $32$ initial conditions as before.
For $T_{m}=T_{\mu}=10$ ($\rho = 1$), the MNAD is reduced from approximately $0.019$, to approximately $0.003$ compared to Latent-LED.
The speed-up, however, is reduced from 60 to 2.
By varying $T_{m} \in \{50,100,200,1000 \}$, Multiscale-LED achieves a trade-off between speed-up and MNAD.

A prediction of the Latent-LED in the inhibitor density is compared against the groundtruth in Figure~\ref{fig:FHN:FHN} G, H, I.
Additional results on the activator density are given in the SI, Section 3B.

\subsection{Kuramoto-Sivashinsky}
\label{sec:kuramoto}
The Kuramoto-Sivashinsky (KS)~\cite{Kuramoto1978,Sivashinsky1977} is a prototypical partial differential equation (PDE) of fourth order that exhibits a very rich range of non-linear phenomena.
In case of high dissipation and small spatial extent $L$ (domain size), the long-term dynamics of KS can be represented on a low dimensional inertial manifold~\cite{robinson1994inertial, linot2020deep}, that attracts all neighboring states at an exponential rate after a transient period.
LED is employed to learn the low order manifold of the effective dynamics in KS.

The one dimensional K-S equation is given by the PDE
\begin{equation}
\begin{split}
&\frac{\partial u}{ \partial t} = - \nu \frac{\partial^4 u}{ \partial x ^4 } - \frac{\partial^2 u}{ \partial x ^2 } - u \frac{\partial u}{ \partial x},\\
\end{split}
\label{eq:kuramoto}
\end{equation}
on the domain $\Omega=[0,L]$ with periodic boundary conditions $u(0,t)=u(L,t)$ and $\nu=1$.
The special case $L=22$ considered in this work, is studied extensively in~\cite{cvitanovic2010state}, and exhibits a structurally stable chaotic attractor, i.e. an inertial manifold where the long-term dynamics lie.
\Cref{eq:kuramoto} is discretized with a grid of size $64$ points, and solved using the fourth-order method for stiff PDEs introduced in~\cite{Kassam2005} with a time-step of $\delta t=2.5 \cdot 10^{-3}$ starting from a random initial condition.
The data are subsampled to $\Delta t=0.25$ (coarse time-step of LED).
$15 \cdot 10^3$ samples are used for training and another $15 \cdot 10^3$ for validation.
For testing purposes, the process is repeated with a different random seed, generating another $15 \cdot 10^3$ samples.

\begin{figure*}[tbhp]
\includegraphics[width=0.9\textwidth]{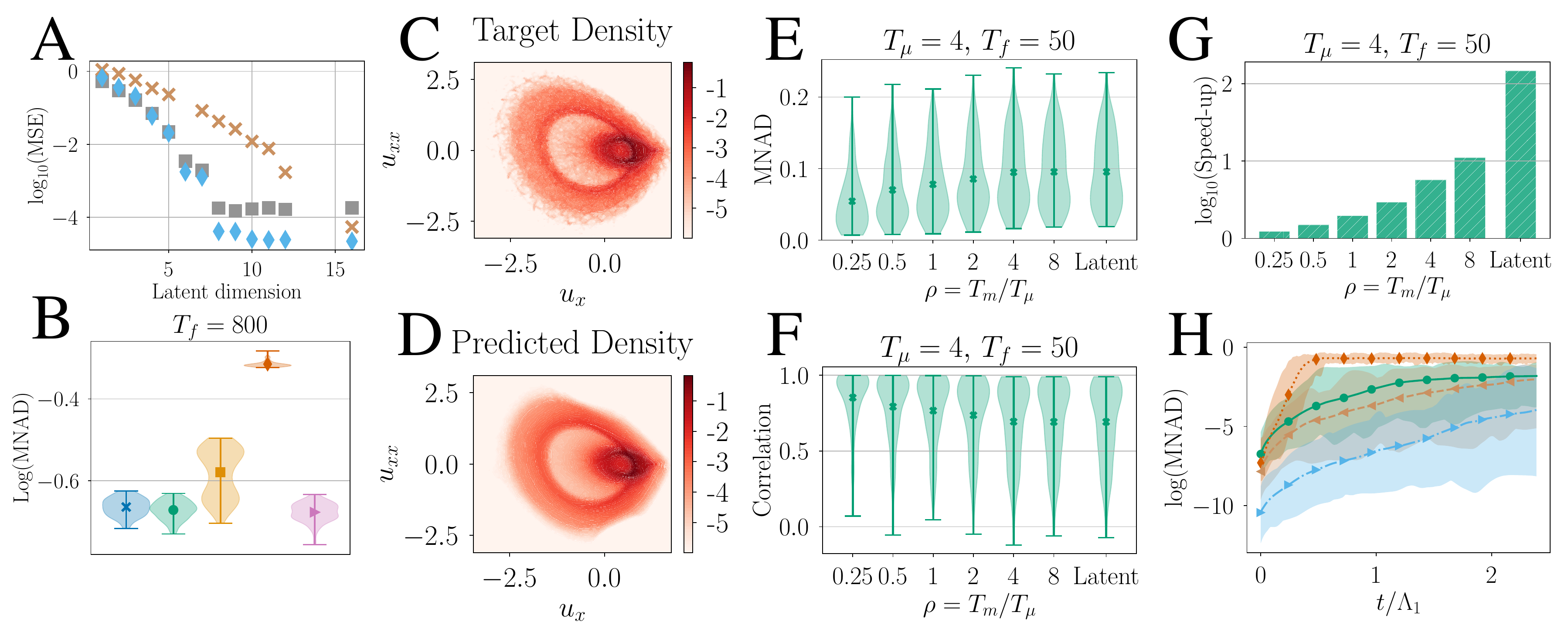}
\caption{
\textbf{A)} Comparison of the reconstruction mean squared error (MSE) in the test data in the FHN dynamics between PCA (\addlegend{L3_cross_brown.pdf}{0.5}), Autoencoder (\addlegend{L3_square_gray.pdf}{0.5}), and Convolutional Autoencoder (\addlegend{L3_diamond_lightblue.pdf}{0.5}) as a function of the latent dimension.
\textbf{B)} Comparison of different macrodynamics propagators (\addlegend{L_cross_blue.pdf}{0.95} AE-LSTM-end2end; \addlegend{L_circle_tirquaz.pdf}{0.95} AE-LSTM; \addlegend{L_rectange_yellow.pdf}{0.95} AE-MLP; \addlegend{L_diamond_orange.pdf}{0.95} AE-RC; \addlegend{L_tiltedtriangle_purple.pdf}{0.95} AE-SINDy) in iterative latent forecasting.
\textbf{C)} The density of values in the $\bm{u}_{x}-\bm{u}_{xx}$ space computed from $100$ reference trajectories that matches closely the \textbf{D)} prediction, illustrating that the LED is able to replicate characteristics of the dynamical system and remain at the attractor, even though propagating coarse dynamics.
\textbf{E)} The MNAD and \textbf{F)} the correlation between LED predictions and the reference data as a function of the multiscale ratio $\rho$.
\textbf{G)} The speed-up of LED w.r.t. $\rho$.
Evolution of the latent state of LED ($T_{\mu}=0$) is up to two orders of magnitude cheaper than the micro scale dynamics.
\textbf{H)} Comparison of CNN-LSTM (LED) (\addlegend{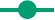}{0.72}), CNN-RC (\addlegend{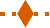}{0.72}), LSTM (\addlegend{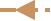}{0.72}), and RC (\addlegend{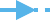}{0.72}) in short-term forecasting of the KS dynamics, time is normalized with the Lyapunov time $T^{\Lambda_1}=1/\Lambda_1 = 20.83$.
}
\label{fig:KS:KS}
\end{figure*}

For the identification of a reasonable latent space dimension, we compare PCA, AEs, and CNNs in terms of the reconstruction MSE in the test data as a function of $d_{\bm{z}}$, plotted in Figure~\ref{fig:KS:KS} A.
MSE is plateauing after $d_{\bm{z}}=8$ indicating arguably the dimensionality of the attractor in agreement with previous studies~\cite{cvitanovic2010state, linot2020deep}, and that the CNN is superior to the AE, while orders of magnitude better than PCA.
For this reason, we employ a CNN with $d_{\bm{z}}=8$ for the autoencoding part of LED.
The hyper-parameters of the networks are tuned based on the MSE on the validation data, reported in SI, Tables 8, 9 with the network training times.
The CNN architecture is provided in SI Table 10, and depicted in SI Figure 12.

In Figure~\ref{fig:KS:KS} B, we compare various propagators in predicting the macro dynamics of LED, starting from $100$ test initial conditions, up to $T_f=800$ ($3200$ time-steps).
We employ a CNN-LSTM trained end-to-end (CNN-LSTM-end2end), a CNN-LSTM where the CNN is pretrained (CNN-LSTM), a multi-layered perceptron (CNN-MLP), Reservoir Computers (CNN-RC)~\cite{pathak2018model, vlachas2020backpropagation}, and the SINDy algorithm (CNN-SINDy)~\cite{brunton2016discovering}.
As a comparison metric, we consider the MNAD, averaged over the $100$ initial conditions.
The warm-up period for all propagators is set to $T_{warm}=60$.
The hyper-parameters (reported on SI Tables 11, 12, 13, along with the training times) are tuned based on the MNAD on the validation data.
While the MLP and RC propagators exhibit large errors, the LSTM, LSTM-end2end, and SINDy show comparable accuracy.
In the following, we consider an LSTM propagator for the LED.

Due to chaoticity of the KS equation, iterative forecasting with LED is challenging, as initial errors propagate exponentially.
In order to assess whether the iterative forecasting with LED leads to reasonable, physical predictions, we plot the density of values in the $\bm{u}_{x}-\bm{u}_{xx}$ space in \Cref{fig:KS:KS} C. The data come from a single long trajectory of size $T_f=8000$ ($32000$ time-steps). 
We observe that LED,~\Cref{fig:KS:KS} D, is able to qualitatively reproduce the density of the simulation.

In~\Cref{fig:KS:KS} E and F we plot the MNAD, and correlation between forecasts of LED and the reference with respect to the multiscale ratio $\rho$.
In~\Cref{fig:KS:KS}  G the speed-up of LED is plotted against $\rho$.
Latent-LED is able to reproduce the long-term ``climate dynamics''~\cite{vlachas2020backpropagation}, and remain at the attractor, while being more than two orders of magnitude faster compared to the micro solver.
As $\rho$ is increased, the error is reduced (correlation increased), at the cost of reduced speed-up.

Finally, in~\Cref{fig:KS:KS} H, we compare the performance of Latent-LED (CNN-LSTM) with previous studies~\cite{pathak2018model, vlachas2020backpropagation}, that forecast directly on the high-dimensional space.
Specifically, the Latent-LED matches the performance of an LSTM (no dimensionality reduction), but shows inferior short-term forecasting ability compared to an RC (no dimensionality reduction) forecasting on the high-dimensional space.
This is expected as the RC and the LSTM have full information of the state.
In turn, when the RC is employed on the latent space of LED as a macro-dynamics propagator, the error grows significantly and the performance is inferior to the CNN-LSTM case.

A forecast of Latent-LED is provided in the SI, Figure 11.

\subsection{Viscous Flow Behind a Cylinder}
\label{sec:navierstokes}

The flow behind a cylinder is a widely studied problem in fluids~\cite{zdravkovich1997flow}, that exhibits a rich range of dynamical phenomena like the transition from laminar to turbulent flow in high Reynolds numbers, and is used as a benchmark for reduced order modeling (ROM) approaches.
The flow behind a cylinder in the two dimensional space is simulated by solving the incompressible Navier-Stokes equations with Brinkman penalization to enforce the no-slip boundary conditions on the surface of the cylinder~\cite{rossinelli2015mrag,Bost2010}.
More details on the simulation are provided in the SI Section 3D.
We consider the application of LED to two Reynolds' numbers $Re \in \{100, 1000\}$.
The definition of $Re$ is provided in the SI Equation (24). 

The flow is simulated in a cluster with $12$ CPU Cores, up to $T=200$, after discarding initial transient.
$250$ time-steps distanced $\Delta t= 0.2$ in time (total time $T=50$) are used for training, $250$ for validation, and the rest for testing purposes.
The vortex sheeding period is $T\approx 2.86$ for $Re=100$, and $T\approx 2.22$ for $Re=1000$.

The state of LED is $\mathbf{s}_t \equiv \{ p, u_x, u_y, \omega \} \in \mathbb{R}^{4 \times 512 \times 1024}$, where $\omega$ is the vorticity field.
For the autoencoding part, LED employs CNNs that take advantage of the spatial correlations.
The architecture of the CNN is given in Table 14 and depicted in Figure 13 in the SI.
The dimension of the latent space is tuned based on the performance on the validation dataset to $d_{\bm{z}}=4$ for $Re=100$ and $d_{\bm{z}}=10$ for $Re=1000$.

The LSTM propagator of LED is benchmarked against SINDy and RC in predicting the dynamics, starting from $10$ initial conditions randomly sampled from the test data for a prediction horizon of $T=20$ (100 time-steps).
The hyper-parameters (reported on SI Tables 15, 16, 17, along with the training times) are tuned based on the MNAD on the validation data.
The logarithm of the MNAD is given in Figure~\ref{fig:cylReComparison} A for $Re=100$ and E for $Re=1000$.
For the $Re=100$ case, the LSTM exhibits lower MNAD and lower variance compared to RC and SINDy. 
For the challenging $Re=1000$ scenario, LSTM and RC exhibit lower MNAD compared to SINDy, with the LSTM being more robust (lower variance).

\begin{figure*}[tbhp]
\includegraphics[width=0.99\textwidth]{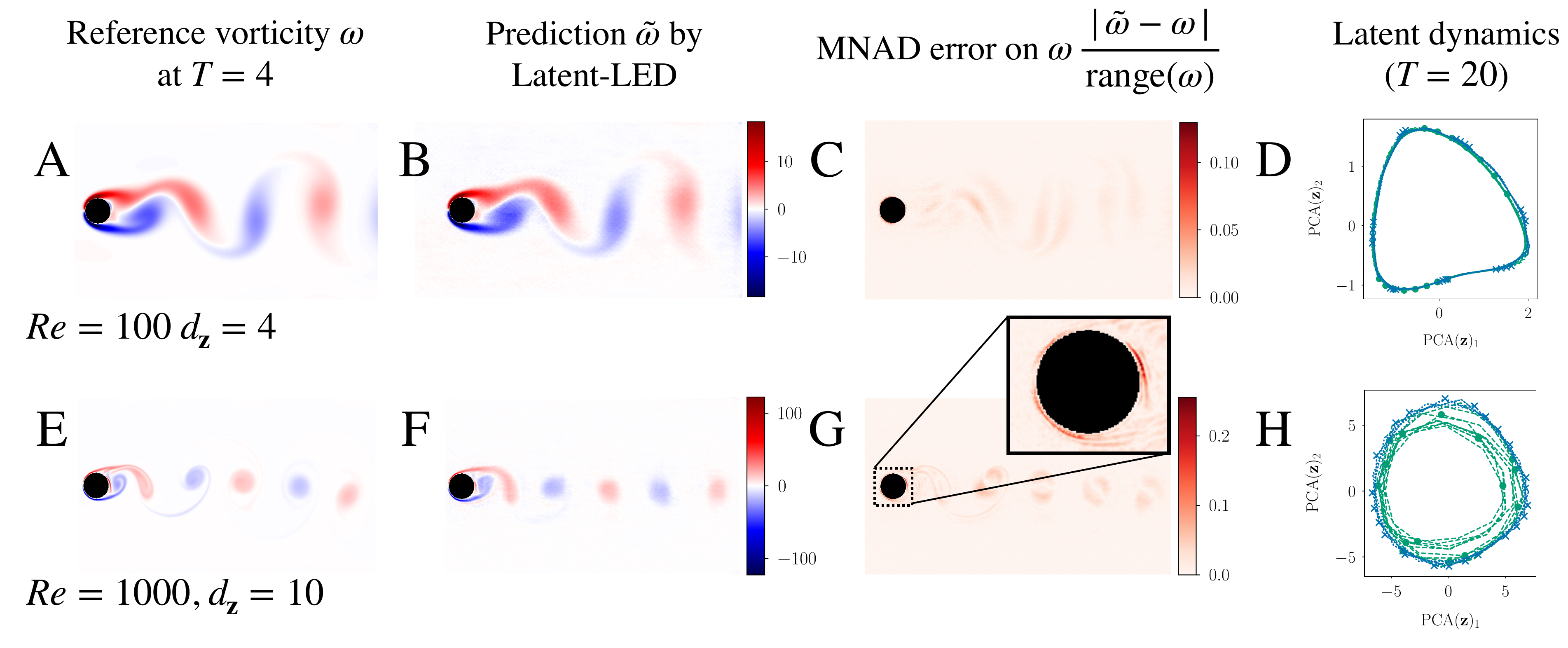}
\caption{
\textbf{A)} The vorticity field $\omega$ at lead time $T=4$ for $Re=100$, and \textbf{E)}  for $Re=1000$.
\textbf{B)} The vorticity field $\tilde{\omega}$ predicted by Latent-LED at final time $T=4$ for $Re=100$, and \textbf{F)}  for $Re=1000$.
\textbf{C)} The MNAD for $Re=100$, and \textbf{G)} for $Re=1000$.
\textbf{F)} The LED latent state dynamics (\addlegend{L2_cross_blue_dotted.pdf}{0.72}) compared against the attractor of the data embedded in the latent space (\addlegend{L2_circle_tirquaz_dashed.pdf}{0.72}).
}
\label{fig:cylReExample}
\end{figure*}

\begin{figure*}[tbhp]
\includegraphics[width=0.99\textwidth]{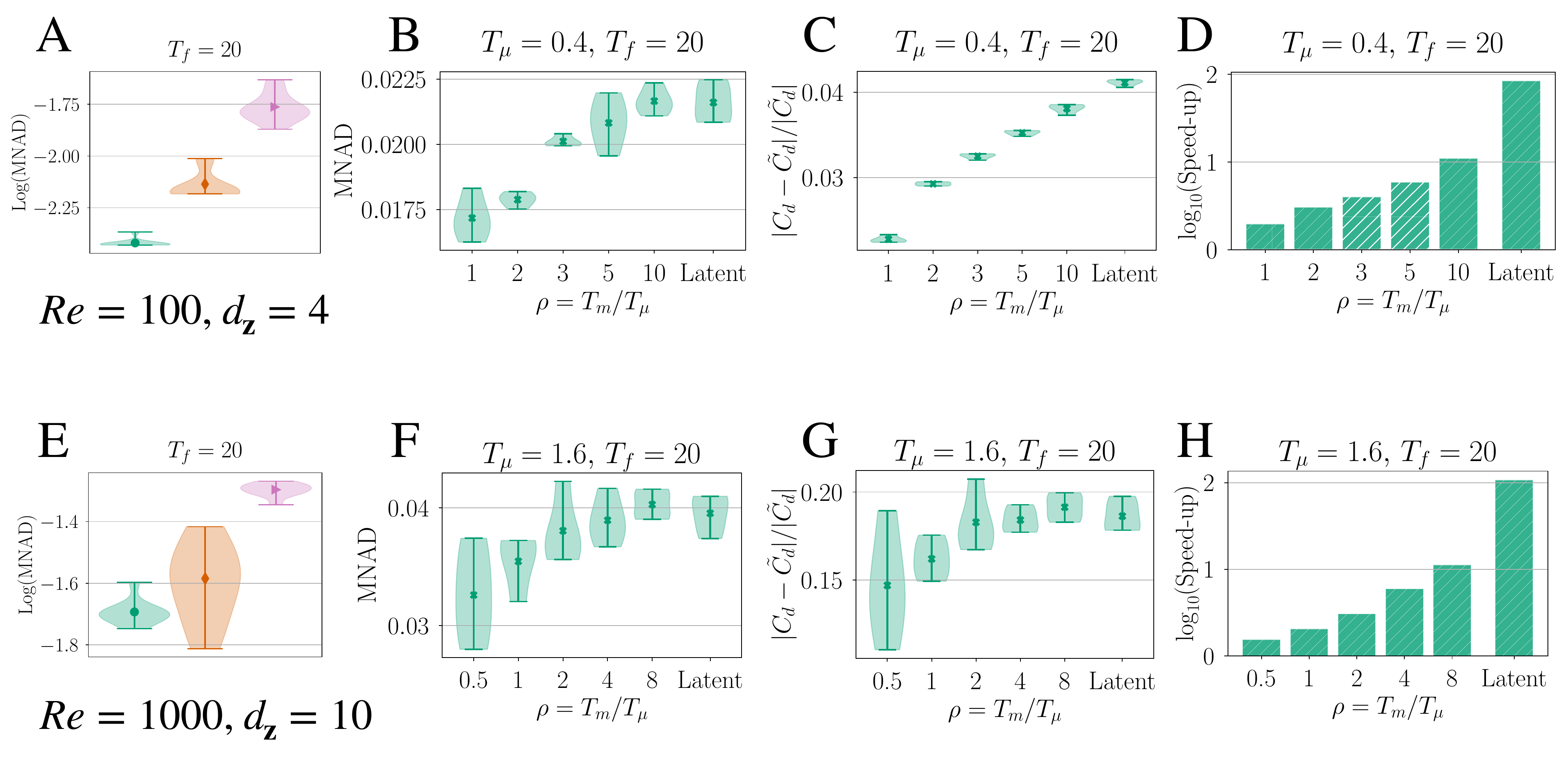}
\caption{
\textbf{A)}  Comparison of different macrodynamics propagators (\addlegend{L_circle_tirquaz.pdf}{0.95} AE-LSTM; \addlegend{L_diamond_orange.pdf}{0.95} AE-RC; \addlegend{L_tiltedtriangle_purple.pdf}{0.95} AE-SINDy) for $Re=100$, and \textbf{E)} for $Re=1000$.
\textbf{B)} The MNAD and \textbf{C)} the relative error on the drag between predictions by LED and the reference data as a function of the multiscale ratio $\rho$, \textbf{F)}, \textbf{G)} the same for $Re=1000$.
\textbf{D)} The speed-up of LED compared to the flow solver w.r.t. $\rho$ for $Re=100$, and \textbf{H)}  for $Re=1000$.
}
\label{fig:cylReComparison}
\end{figure*}

A prediction of the vorticity $\omega$ by Latent-LED at lead time $T=4$ is given in Figure~\ref{fig:cylReExample}.
LED captures the flow for both $Re \in \{100,1000 \}$.
The error concentrates mostly around the cylinder, rendering the accurate prediction of the drag coefficient challenging.
In Figure~\ref{fig:cylReExample} D and H, the latent space of Latent-LED is compared with the transformation of the data to the latent space.
The predictions stay close to the attractor even for very large horizon ($T=20$).
The Strouhal number $\operatorname{St}$ (defined in the SI Equation (23)) describes the periodic vortex shedding at the wake of the cylinder.
By estimating the dominant frequency of the latent state using a Fourier analysis, we find that LED reproduces exactly the $\operatorname{St}$ of the system dynamics for both $Re\in \{100, 1000\}$ cases.

In the $Re=100$ case, Latent-LED recovers a periodic non-linear mode in the latent space, and can forecast the dynamics accurately, as illustrated in Figure~\ref{fig:cylReExample}.
In this case, approaches based on the Galerkin method or dynamic mode decomposition (DMD), construct ROM with six to eight degrees of freedom~\cite{taira2020modal} that capture the most energetic spatiotemporal modes.
In contrast, the latent space of LED in the $Re=100$ case has a dimensionality of $d_{\bm{z}}=4$.
In the challenging $Re=1000$ scenario, LED with $d_{\bm{z}}=10$ can capture accurately the characteristic vortex street, and long-term dynamics.
We note that, to the best of our knowledge, ROMs for flows past a  cylinder have been so far limited to laminar periodic flows in the order of $Re=100$ while this study advances the state of the art by one order of magnitude.

Starting from $4$ initial conditions randomly sampled from the test data, six LED variants (
Latent-LED, Multiscale-LED with $T_{\mu}=0.4, T_m \in \{0.4, 0.8, 1.2, 2, 4\}$ for $Re=100$, and 
Latent-LED, Multiscale-LED with $T_{\mu}=1.6, T_m \in \{0.8, 1.6, 3.2, 6.4, 12.8\}$ for $Re=1000$
) are tested on predicting the dynamics of the flow up to $T_f=20$, after $T_{warm}=2$.
The MNAD is plotted in Figure~\ref{fig:cylReComparison} B for $Re=100$, and F for $Re=1000$.
The speed-up is plotted in Figure~\ref{fig:cylReComparison} D for $Re=100$, and H for $Re=1000$.
The Latent-LED is two orders of magnitude faster than the flow solver, while exhibiting MNAD errors of $0.02$ and $0.04$ for $Re=100$, and $Re=1000$ respectively.
By alternating between macro and micro, the error is reduced, at the cost of decreased speed-up.

In Figure~\ref{fig:cylReComparison} C and G, the relative error on the drag coefficient $C_d$ (defined in SI, Equation (28)) is plotted as a function of the multiscale ratio $\rho$.
Latent-LED exhibits a relative error of $0.04$ that is reduced to approximately $0.02$ for $\rho=1$.
For $Re=1000$, as we observe in~\Cref{fig:cylReExample}, the prediction error of LED concentrates around the cylinder which leads to an inaccurate computation of the drag.
Even though Multiscale-LED is reducing this error, it still remains on the order of $0.15$.

\section{Discussion}

We have presented a novel framework for learning the effective dynamics (LED) and accelerate the simulations of multiscale (stochastic or deterministic) complex dynamical systems.
Our work relies on augmenting the Equation-Free formalism with state of the art ML methods.

The LED framework is tested on a number of benchmark problems.
In systems where evolving the high-dimensional state dynamics is computationally expensive, LED accelerates the simulation by propagating on the latent space and upscaling to the high-dimensional states with the probabilistic, generative mixture density, or deterministic convolutional, decoder.
This comes at the cost of training the networks, a process that is performed once, offline.
The trained model can be used to forecast the dynamics starting from any arbitrary initial condition.

The efficiency of LED was evaluated in forecasting the FitzHugh-Nagumo equation dynamics achieving an order of magnitude lower approximation error compared to other Equation-Free approaches while being two orders of magnitude faster than the Lattice Boltzmann solver.
We demonstrated that the proposed framework identifies the effective dynamics of the Kuramoto-Sivashinsky equation with $L=22$, capturing the long-term behavior (``climate dynamics''), achieving a speed-up of $S\approx 100$.
Furthermore, LED captures accurately the long-term dynamics of a flow behind a cylinder in $Re=100$ and $Re=1000$, while being two orders of magnitude faster than a flow solver.
In the SI, we demonstrate that LED can unravel and forecast the stochastic collective dynamics of $1000$ particles following Brownian motion subject to advection and diffusion in the three dimensional space (SI, Section 3A).
In our recent work~\cite{vlachas2021accelerated} (briefly described in SI, Section 3E), we show that LED can be applied to learn the stochastic dynamics of molecular systems.
We note that the present method is readily applicable to all problems where Equation-Free, HMM, and FLAVOR methodologies have been applied.

In summary, LED identifies and propagates the effective dynamics of dynamical systems with multiple spatiotemporal scales providing significant computational savings.
Moreover, LED provides a systematic way of trading between speed-up and accuracy for a multiscale system by switching between propagation of the latent dynamics, and evolution of the original equations, iteratively correcting the statistical error at the cost of reduced speed-up.

The LED does not presently contain any mechanism to decide when to upscale the latent space dynamics. 
This is an active area of investigations.
We do not expect LED to generalize to dynamical regions drastically different from those represented in the training data.
Further research efforts will address this issue by adapting the training procedure.

The present methodology can be deployed both in problems described by first principles as well as for problems where only data are available for either the macro or microscale descriptions of the system.
LED creates unique algorithmic alloys between data driven and first principles models and opens new horizons for the accurate and efficient prediction of complex multiscale systems.

\begin{acknowledgments}
The authors would like to thank Nikolaos Kallikounis (ETH Zurich) for helpful discussions on the Lattice Boltzmann method, Pascal Weber and Michalis Chatzimanolakis (ETH Zurich) for help with the simulations of the flow behind a cylinder, and Yannis Kevrekidis (Johns Hopkins University),Kostas Spiliotis (University of Rostock), for providing code to reproduce data for the FHN equation.
The authors acknowledge the support of the Swiss National Supercomputing Centre (CSCS) providing the necessary computational resources under Projects s929.
\end{acknowledgments}

\section*{Author contribution}
P.K. conceived the project;
P.R.V., G.A., C.U., and P.K. designed and performed research;
P.R.V., and G.A. contributed new analytic tools;
P.R.V., G.A., and P.K. analyzed data;
and P.R.V., G.A., and P.K. wrote the paper.

\section*{Author Declaration}
The authors declare no conflict of interest.

\section*{Data and Code Availability}

All code and data for the analysis associated with the current submission will become readily available upon publication in the following link: \href{https://github.com/pvlachas/LearningEffectiveDynamics}{https://github.com/pvlachas/LearningEffectiveDynamics}.

\bibliographystyle{naturemag}

\bibliography{references}

\clearpage

\appendix

\counterwithout{equation}{section}

\renewcommand\thesection{\Roman{section}}
\renewcommand\thesubsection{\Alph{subsection}}

\setcounter{figure}{0}    
\setcounter{table}{0}
\setcounter{equation}{0}

\section{Methods}
\label{sec:methods}

The framework to learn and propagate the effective dynamics (LED) of complex systems is composed of the models described in the following.




\subsection{Autoencoders (AE)}
\label{sec:autoencoders}

\begin{figure*}[tbhp]
\centering
\begin{subfigure}[tbhp]{0.4\textwidth}
\centering
\includegraphics[width=1.0\textwidth]{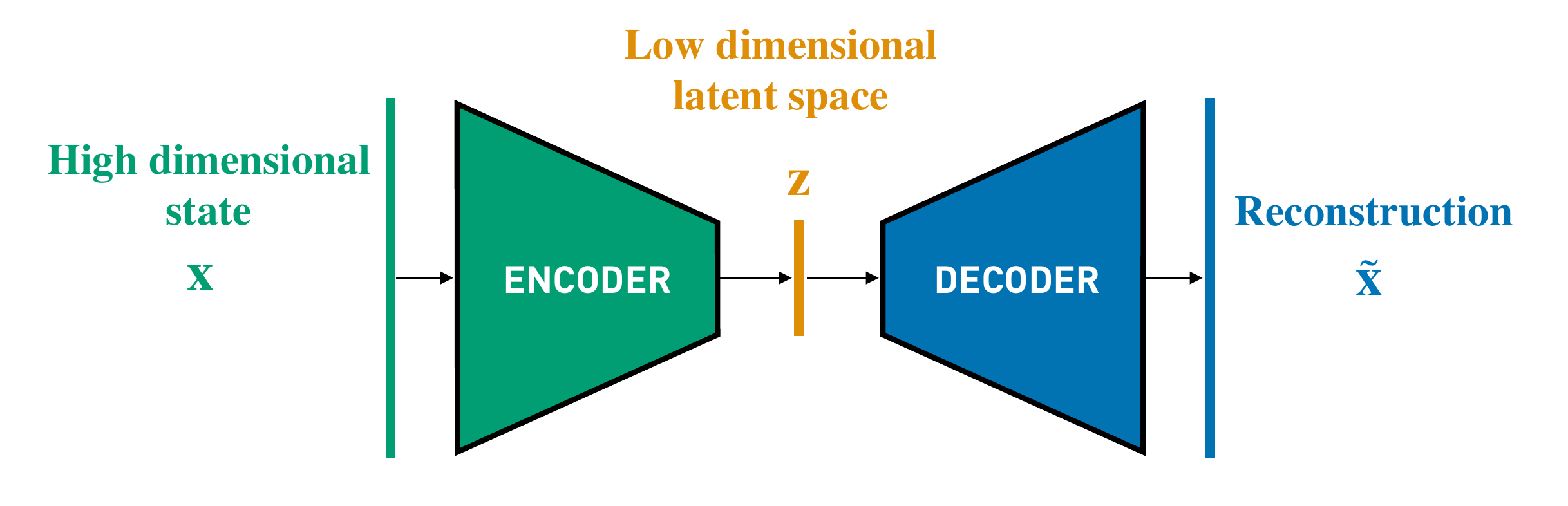}
\caption{Classical Autoencoder (AE)}
\label{fig:autoencoder:autoencoder}
\end{subfigure}
\hfill
\begin{subfigure}[tbhp]{0.5\textwidth}
\centering
\includegraphics[width=1.0\textwidth]{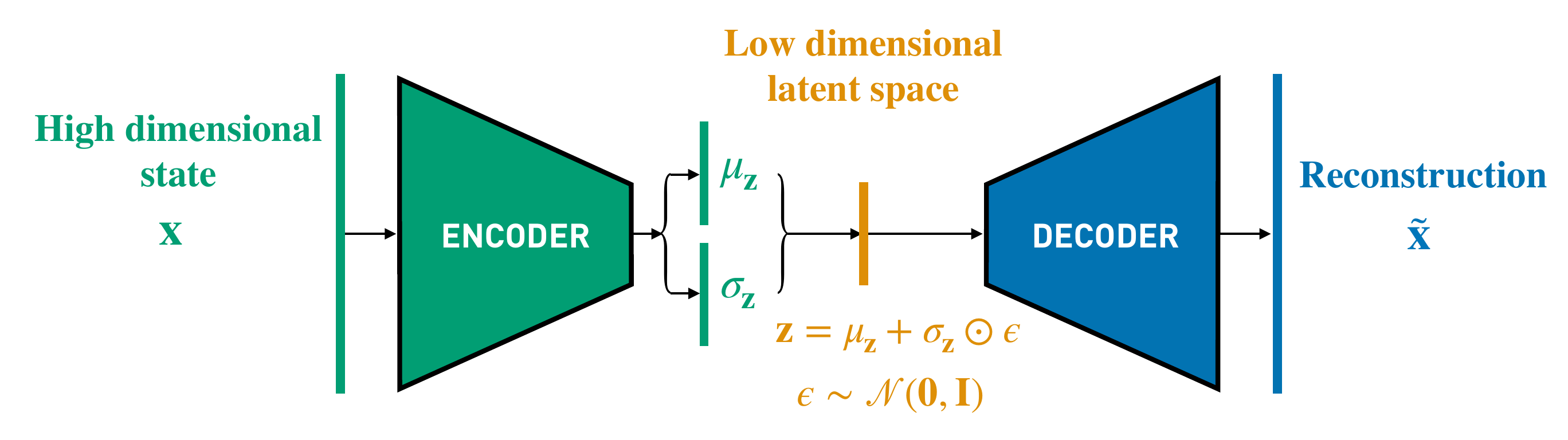}
\caption{Variational Autoencoder (VAE)}
\label{fig:autoencoder:variationalAutoencoder}
\end{subfigure}
\caption{
\textbf{(a)} A schematic diagram of a classical Autoencoder (AE). A high-dimensional state $\bmx$ is mapped to a low dimensional feature space $\bmz$ by applying the encoder transformation through multiple fully connected layers.
The low dimensional feature space $\bmz$ is expanded in the original space by the decoder.
The autoencoder is trained with the loss $\calL=||\bmx - \bmtx||^2$, so that the input can be reconstructed as faithfully as possible at the decoder output.
\textbf{(b)} A schematic diagram of a Variational Autoencoder (VAE). 
Instead of modeling the latent space deterministically, the encoder outputs a mean latent representation $\bm{\mu_{z}}$, along with the associated uncertainty $\sigma_{\bm{z}}$.
The latent space $\bmz$ is sampled from a normal distribution $\bmz \sim \calN( \cdot | \bm{\mu_{z}},  \sigma_{\bm{z}} \bm{I} )$, with diagonal covariance matrix.
}
\label{fig:autoencoder}
\end{figure*}

Classical autoencoders are non-linear neural networks that map an input to a low dimensional latent space and then decode it to the original dimension at the output, trained to minimize the reconstruction loss $\calL=|\bmx-\bmtx|^2$.
They were proposed in as a non-linear alternative to Principal Component Analysis (PCA).
An autoencoder is depicted in \Cref{fig:autoencoder:autoencoder}.

In this work, we employ  feed-forward AEs to identify the coarse representation of the Fitz-Hugh Nagumo equation, and the Kuramoto-Sivashinsky equation.

\subsection{Variational Autoencoders (VAE)}
\label{sec:vae}

Research efforts on generative modeling led to the development of Variational Autoencoders (VAEs).
The VAE similar to AE is composed by an encoder and a decoder.
The encoder neural network, instead of mapping the input $\bmx$ deterministically to a reduced order latent space $\bmz$, produces a distribution $q(\bmz | \bmx ; \bmw_{q})$ over the latent representation $\bmz$, where $\bmw_{q}$ is the parametrization of the distribution given by the output of the encoder $\bmw_{q} = \encoder (\bmx) $.
In most practical applications,  the distribution $q(\bmz | \bmx ;\bmw_{q})$ is modeled as a factorized Gaussian, implying that $\bmw_{q}$ is composed of the mean, and the diagonal elements of the covariance matrix.
The decoder maps a sampled latent representation to an output $\bmtx = \decoder (\bmz)$.
By sampling the latent distribution $q(\bmz | \bmx ; \bmw_{q})$, for a fixed input $\bmx$, the autoencoder can generate samples from the probability distribution over $\bmtx$ at the decoder output.
The network is trained to maximize the log-likelihood of reproducing the input at the output, while minimizing the Kullback-Leibler divergence between the encoder distribution $q(\bmz | \bmx ; \bmw_{q})$ and a prior distribution, e.g. $\calN(0, \bm{I})$.
VAEs are esentially regularizing the training of AE by adding the Gaussian noise in the latent representation.
In this work, a Gaussian latent distribution with diagonal covariance matrix is considered, i.e.,
\begin{equation}
q(\bmz \, | \, \bmx ; \bm{\mu}_{\bmz}, \bm{\sigma}_{\bmz} ) = \calN \big( \bmz \, | \, \bm{\mu}_{\bmz}(\bmx),  \operatorname{diag}( \sigma_{\bmz}(\bmx) )  \big ),
\end{equation}
where $ \bmw_{q}  = ( \bm{\mu}_{\bmz}, \sigma_{\bmz} )$ and the mean latent representation $\bm{\mu}_{\bmz}$ and the variance $\sigma_{\bmz}$ vectors are the outputs of the encoder neural network $\encoder (\bmx) $.
The latent representation is then sampled from $\bmz \sim \calN(\bm{\mu}_{\bmz}, \operatorname{diag}( \sigma_{\bmz} ) )$.
The decoder receives as an input the sample, and outputs the reconstruction $\bmtx$.
A VAE is depicted in \Cref{fig:autoencoder:variationalAutoencoder}.

A preliminary study, benchmarking VAEs against feedforward AEs (and Convolutional AE described later) in the FitzHugh-Nagumo equation, and the Kuramoto-Sivashinsky equation, showed no significant advantages for the cases considered in this work over feed-forward AEs.
They are, however, part of the LED framework, and may be useful in other applications.

\subsection{Convolutional Neural Networks}
\label{sec:cnns}

Convolutional neural networks (CNNs) are tailored to process image data with spatial correlations.
Each layer of a CNN is processing a multidimensional input (with a channel axis, and some spatial axes) by applying a convolutional kernel or filter that slides along the input spatial axes.
In other words, CNNs take into account of the structure in the data in their architecture, which is a form of a geometric prior.
In this work, CNN layers are used in the Autoencoder, by introducing a bottleneck layer, reducing the dimensionality.
Other dimensionality reduction techniques, like AEs, Principal Component Analysis (PCA), or Diffusion maps (DiffMaps), that are based on vectorization of input field data, do not take into account the structure of the data, i.e. when an input field is shifted by a pixel, the vectorized version will differ a lot, while the convoluted image will not.

In this work, we employ Autoencoding CNNs (and compare them with feed-forward AEs) to identify the coarse representation of the FitzHugh-Nagumo equation, the Kuramoto-Sivashinsky equation, and the incompressible Navier-Stokes flow behind a cylinder at $Re\in \{100,1000 \}$.


\subsection{Permutation Invariance}
\label{sec:perminv}

\begin{figure}[tbhp]
\centering
\includegraphics[width=0.975\textwidth]{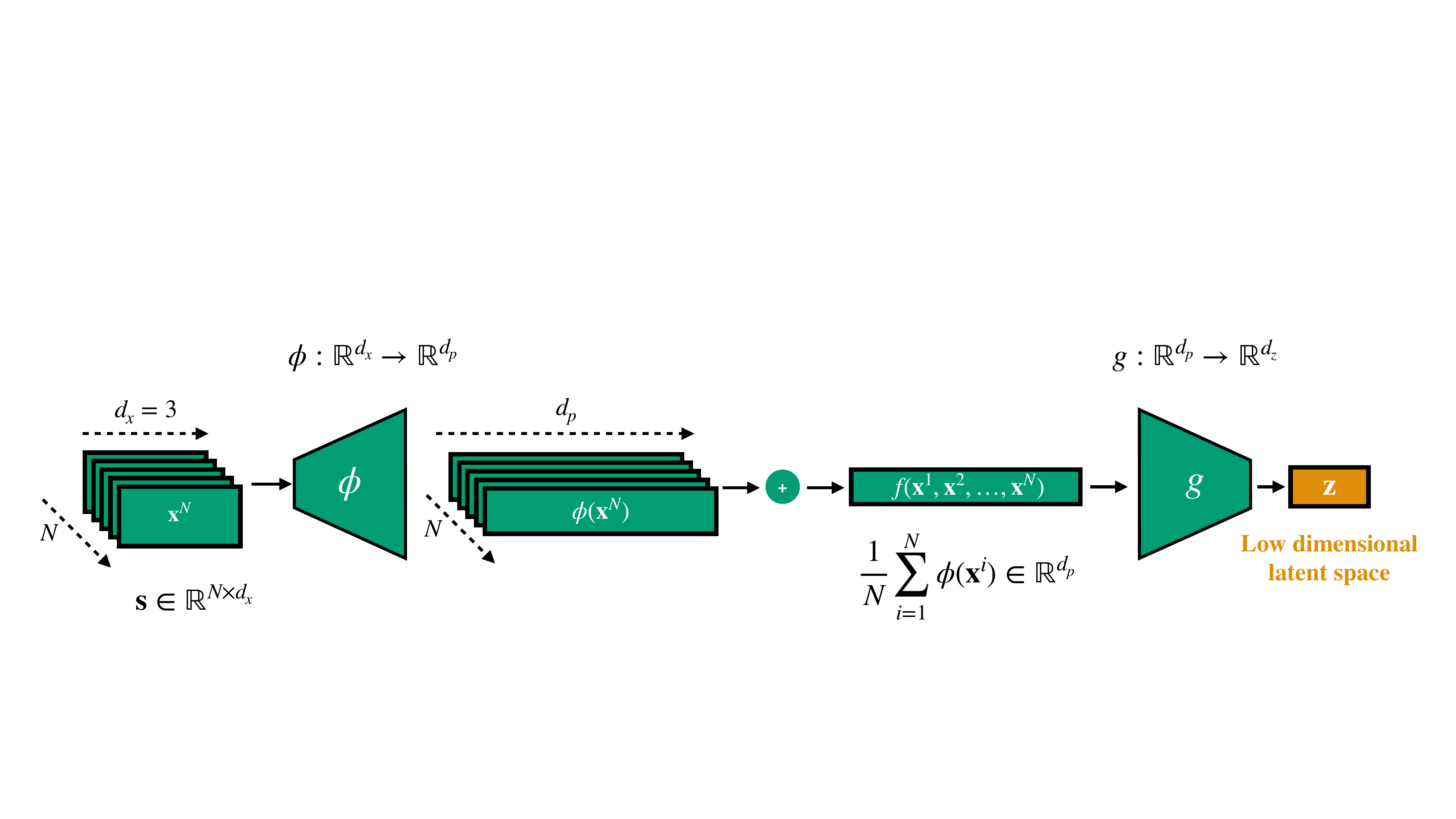}
\label{fig:permInvLayer}
\caption{
Illustration of the permutation invariant encoder.
The input of the network is composed of $N$ atomic states that are permutation invariant, e.g. positions $\{\bmx^1, \dots, \bmx^N \}$ of $N$ particles in a particle simulation, each one with dimension $d_{\bmx}$, i.e. $\bmx^i \in \mathbb{R}^{d_{\bmx}}, \, \forall i \in \{1,\dots,N\}$.
A transformation $\phi(\cdot): \mathbb{R}^{d_{\bmx}} \to \mathbb{R}^{d_p}$ is applied to each atomic state separately, mapping to a high-dimensional latent feature space.
The mean of these latent representations of the atomic states is computed, leading to a singe latent feature that is permutation invariant with respect to the input.
The final layer of the encoder maps the high-dimensional feature to a low dimensional representation $\bmz$, which is again permutation invariant with respect to the input, representing the encoding of the global state.
}
\end{figure}

Physical systems may satisfy specific properties like energy conservation, translation invariance, permutation invariance, etc.
In order to build data-driven models that accurately reproduce the statistical behavior of such systems, these properties should be embedded in the model.
In this section, the dynamics of particles of the same kind are modeled with a permutation invariance layer.
This is useful in simulations of molecules, i.e. molecular dynamics, where the state of the system is described by a configuration of particles, and any permutation of these particles corresponds to the same configuration.
Permutation invariance is handled here with a sum decomposition of a feature space.
The exact procedure is depicted in~\Cref{fig:permInvLayer}.

Assume that the state of a dynamical system $\bms$ is composed of $N$ particles of the same kind, each one having specific properties or features with dimensionality $d_{\bmx}$, e.g. position, velocity, etc.
The features of a single particle are given by the state $\bmx \in \mathbb{R}^{d_{\bmx}}$ of the particle.
Raw data is provided as an input to the network, i.e. the features of all particles, stacked together in a matrix $\bms \in \mathbb{R}^{N \times d_{\bmx}}$.
A permutation of two particles represents in essence the same configuration and should be mapped to the same latent representation.
This is achieved with a permutation invariant layer that first applies a non-linear transformation $\phi:\mathbb{R}^{d_{\bmx}} \to \mathbb{R}^{d_p}$ mapping each particles' features to a high-dimensional latent representation of dimension $d_p$.
This mapping is applied to all particles independently leading to $N$ such latent vectors.
The mean of these vectors is taken to construct the representation of the configuration.
The representation $\frac{1}{N} \sum_{i=1}^N \phi(\bmx^i)$ is finally fed to a final layer reducing the dimensionality to a low-order representation $\bmz \in \mathbb{R}^{d_{\bmz}}$, with $d_{\bmz} \ll d_p, N$.
This is achieved by the mapping $g:\mathbb{R}^{d_p} \to \mathbb{R}^{d_{\bmz}} $.
In this work, the permutation invariance layer is utilized in the modeling of the collective dynamics of a group of particles whose movement is governed by the advection-diffusion equation in the one and three dimensional space.
Both mappings $g$ and $\phi$ are implemented with neural networks, having $3$ layers of $50$ hidden units each, and $\tanh$ activations.

\subsection{Mixture Density Decoder}
\label{sec:mdn}

Mixture density networks (MDNs) \cite{bishop1994mixture} are powerful neural networks that can model non-Gaussian, multi-modal data distributions.
The outputs of MDNs are parameters of a mixture density model (mixture of probability density functions).
The most generic choice of the mixture component distribution, is the Gaussian distribution.
Gaussian MDNs are widely deployed in machine learning applications to model structured dynamic environments, i.e. (video) games.
The effectiveness of MDNs, however, in modeling physical systems remains unexplored.

In physical systems, the state may be bounded.
In this case, the choice of a Gaussian MDN is problematic due to its unbounded support.
To make matters worse, most applications of Gaussian MDNs when modeling random vectors do not consider the interdependence between the vector variables, i.e. the covariance matrix of the Gaussian mixture components is diagonal, in an attempt to reduce their computational complexity.
Arguably in the applications where they were successful, modeling this interdependence was not imperative.
In contrast, in physical systems the variables of a state might be very strongly dependent on each other.
In order to cope with these problems, the following approach is considered:
Firstly, an auxiliary vector variable is considered $\bmv$ along with its distribution $p( \bmv | \bmz )$.
$\bmv \in \mathbb{R}^{d_{\bmx}}$ has the same dimensionality $d_{\bmx}$ as the high-dimensional state (input/output of the autoencoder).
The distribution is modeled as a mixture of $K$ \textbf{multivariate} normal distributions
\begin{equation}
p(  \bmv | \bmz) = \sum_{k=1}^K \pi^{k}(\bmz) \, \calN  \bigg( \, \bm{\mu}_{ \bmv }^k(\bmz), \Sigma_{\bmv}^k(\bmz) \, \bigg),
\end{equation}
The multivariate normal distribution is parametrised in terms of a mean vector $\bm{\mu}_{ \bmv }^k$, a positive definitive covariance matrix $\Sigma_{\bmv}^k$, and the mixing coefficients $\pi^{k}$  which are functions of $\bmz$.
The covariance matrix is parametrised by a lower-triangular matrix $L_{\bmv}^k$ with positive-valued diagonal entries, such that $\Sigma_{\bmv}^k=L_{\bmv}^k L_{\bmv}^{k \, T} \in \mathbb{R}^{d_{\bmx} \times d_{\bmx}}$
(This triangular matrix can be recovered by Cholesky factorization of the positive definite $\Sigma_{\bmv}^k$).
The functional forms of $\pi^{k}(\bmz) \in \mathbb{R}$,  $\bm{\mu}_{ \bmv } (\bmz) \in \mathbb{R}^{d_{\bmx}}$, and the $n (n+1)/2$ entries of $L_{\bmv}^k$ are neural networks, their values are given by the outputs of the decoder for all mixture components $k \in \{1,\dots, K \}$, i.e. $ \bmw_{\calD} = \decoder (\bmz) = \{ \pi^{k}, \bm{\mu}_{ \bmv }^k, L_{\bmv}^k \}_{1, \dots, K} $.
The positivity of the diagonal elements of $L_{\bmv}^k$ is ensured by a \textbf{softplus} activation function
\begin{equation}
f(x)=\ln(1+ \exp(x))
\end{equation}
in the respective outputs of the decoder.
The mixing coefficients satisfy $ 0 \leq \pi^k < 1$ and $\sum_{k=1}^K \pi^k=1$.
To ensure these conditions, the respective outputs of the decoder are passed through a \textbf{softmax} activation
\begin{equation}
\sigma(\bmx)_i = \frac{e^{\bmx_i}}{\sum_i e^{\bmx_i}}.
\end{equation}
The rest (non-diagonal elements and mean vector) of the decoder outputs have linear activations, so no restriction in their sign.
In total, the decoder output is composed of $K(n-1)n/2 + Kn$ single valued outputs with linear activation for the non-diagonal elements of $L_{\bmv}^k$ and the mean vectors $\bm{\mu}_{\bmv}^k$, and $Kn$ positive outputs with softplus activation for the diagonal of $L_{\bmv}^k$, and $K$ outputs with softmax activation for the mixing coefficients.

\begin{figure*}[tbhp]
\centering
\includegraphics[width=0.75\textwidth]{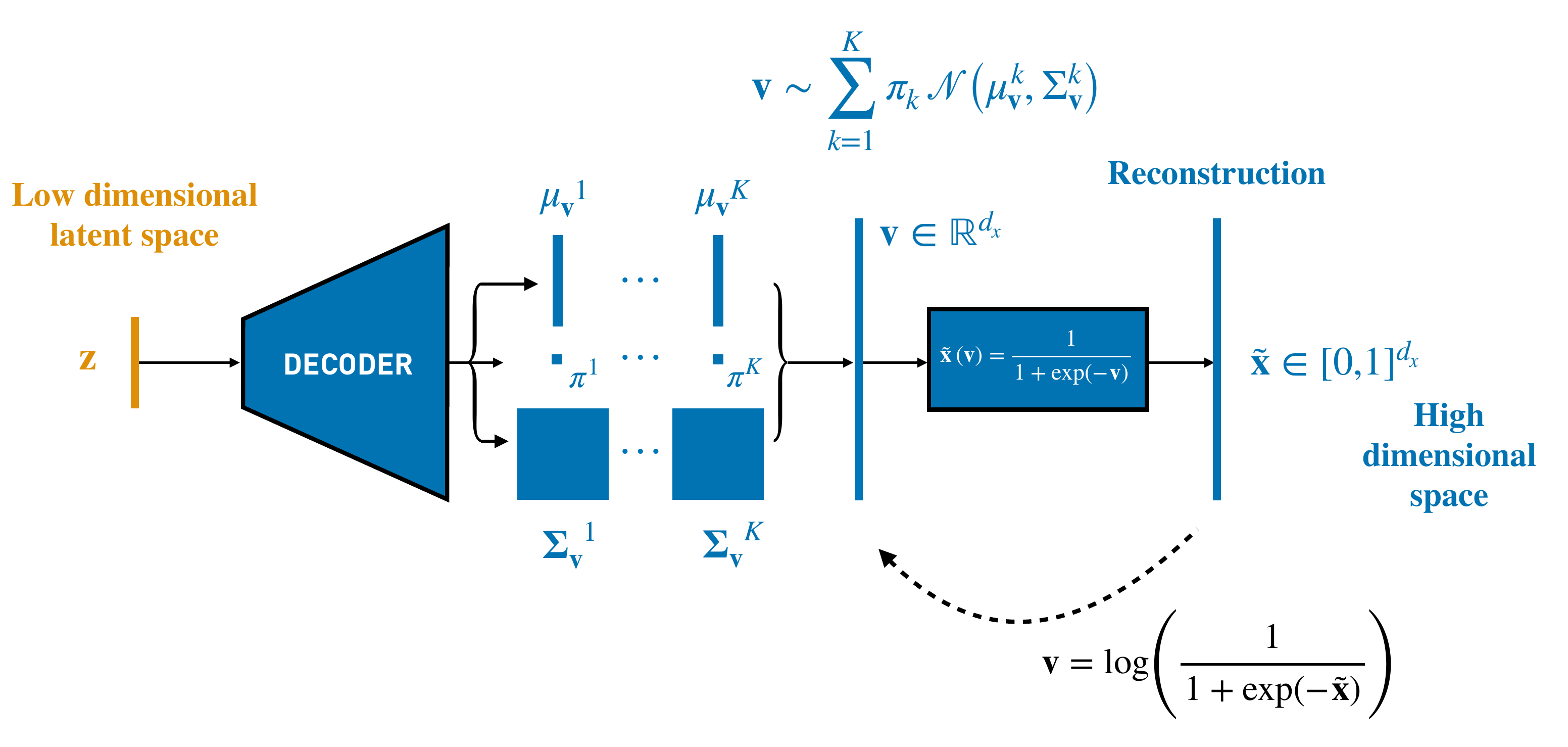}
\caption{
A mixture density network modeling the probability density $p(\bmtx | \bmz)$, with bounded $\bmtx$.
The decoder maps the latent state $\bmz$ to the parameters of a mixture model on the latent vector $\bmv \in \mathbb{R}^{d_{\bmx}}$, which are the mixing coefficients $\pi^k \in \mathbb{R}$, mean vectors $\mathbf{\mu}^{k}_{v} \in \mathbb{R}^{d_{\bmx}}$, and a lower-triangular matrix $L_{\bmv}^k\in \mathbb{R}^{d_{\bmx} \times d_{\bmx}}$ with positive-valued diagonal entries.
From the latter, the covariance matrix is derived from $\Sigma_{\bmv}^k=L_{\bmv}^k L_{\bmv}^{k \, T} $ which is positive definite by construction.
The mixture models the probability distribution of the latent state $p(\bmv | \bmz)$.
The targets, however, used to train the network in a supervised way are defined on the reconstruction $\bmtx$.
The targets are scaled to $\bmtx \in [0,1]^{d_{\bmx}}$, and then transformed to targets for $\bmv$ using the inverse of the softplus activation.
The MDN autoencoder is trained to maximize the likelihood $p(\bmv | \bmz)$ of the transformed data $\bmv$.
}
\label{fig:mdnDecoder}
\end{figure*}

MD networks are employed in stochastic systems (e.g. molecular dynamics).
In the following, we assume that the high-dimensional state of the molecular system is described by $\bm{s}_t \in \mathbb{R}^{d_{\bm{s} }}$.
The decoder $\mathcal{D}^{\bm{w}_{ \mathcal{D}}}$ is modeled with an MD network.
The MD approximates the probability distribution of the state $\bm{\tilde{s}}_t \sim p( \cdot; \bm{w}_{\operatorname{MD}})$, where $\bm{w}_{\operatorname{MD}} = \mathcal{D}^{\bm{w}_{ \mathcal{D}}}(\bm{z}_t)$ is the output of the decoder that parametrizes the distribution.
The optimal parameters of the MD autoencoder are identified by maximizing the log-likelihood of the reconstruction,
\begin{gather*}
\bm{w}_{\mathcal{E}}^{\star}, \bm{w}_{\mathcal{D}}^{\star} = \underset{\bm{w}_{\mathcal{E}}, \bm{w}_{\mathcal{D}}} {\operatorname{argmax}}
\,\log\,p \big( \bm{s}_t ; \bm{w}_{\operatorname{MD}} \big), \\
\textrm{where}\quad
\bm{w}_{\operatorname{MD}} = \mathcal{D}^{\bm{w}_{ \mathcal{D}}}(\bm{z}_t) =
\mathcal{D}^{\bm{w}_{ \mathcal{D}}} \big(
\mathcal{E}^{\bm{w}_{ \mathcal{E}}}(\bm{s}_t)
\big ).
\end{gather*}


\subsection{Long Short-Term Memory Recurrent Neural Networks  (LSTM-RNNs)}
\label{sec:rnns}

In the low order manifold (coarse, latent state), a Recurrent Neural Network (RNN) is utilized to capture the non-linear, non-markovian dynamics.
The forecasting rule of the RNN is given by
\begin{equation}
\bmh_{t} =
\calH^{
\bmw_{\calH}}
\big(
\bmz_t, \bmh_{t-\Delta t}
\big), 
\quad 
\bmtz_{t+\Delta t} =
\calR^{
\bmw_{\calR}
}
\big(
\bmh_t
\big), 
\end{equation}
where $\bmw_{\calH}$ and $\bmw_{\calR}$ are the trainable parameters of the network, $\bmh_t \in \mathbb{R}^{d_{\bmh}}$ is an internal hidden memory state, and $\bmtz_{t+\Delta t} $ is a prediction of the latent state.
The RNN is trained to minimize the forecasting loss $||\bmtz_{t+\Delta t} - \bmz_{t+\Delta t}||_2^2$, which can be written as
\begin{gather}
||\bmtz_{t+\Delta t} - \bmz_{t+\Delta t}||_2^2
= 
||
\calR^{
\bmw_{\calR}
}
\big(
\bmh_t
\big)
- \bmz_{t+\Delta t}
||_2^2
= 
||
\calR^{
\bmw_{\calR}
}
\big(
\calH^{
\bmw_{\calH}}
\big(
\bmz_t, \bmh_{t-\Delta t}
\big)
\big)
- \bmz_{t+\Delta t}
||_2^2
.
\end{gather}
This leads to
\begin{gather}
\bmw_{\calH}, \bmw_{\calR} = \underset{ \bmw_{\calH}, \bmw_{\calR} }{\operatorname{argmin}}
||
\calR^{
\bmw_{\calR}
}
\big(
\calH^{
\bmw_{\calH}}
\big(
\bmz_t, \bmh_{t-\Delta t}
\big)
\big)
- \bmz_{t+\Delta t}
||_2^2
.
\end{gather}
The RNNs are trained with Backpropagation through time (BPTT)~\cite{Werbos1988}.
The mappings $\calH^{
\bmw_{\calH}}$
and
$\calR^{
\bmw_{\calR}
}$, considered in this work take the form of the long short-term memory (LSTM)~\cite{Hochreiter1997} cell.
The output mapping is given by a linear transformation, i.e.
\begin{gather}
\bmtz_{t+\Delta t} =
W_{\bmz, \bmh}
\bmh_t
, 
\end{gather}
where $W_{\bmz, \bmh} \in \mathbb{R}^{d_{\bmz} \times d_{\bmh}}$.
As a consequence, the set of trainable weights of the hidden-to-output mapping is just one matrix $\bmw_{\calR}=W_{\bmz, \bmh}\in \mathbb{R}^{d_{\bmz} \times d_{\bmh}}$.

The LSTM possesses two hidden states, a cell state $\bmc$ and an internal memory state $\bmh$.
The hidden-to-hidden mapping
\begin{equation}
\bmh_{t}, \bmc_{t} =
\calH^{
\bmw_{\calH}}
\big(
\bmz_t, \bmh_{t-\Delta t}, \bmc_{t-\Delta t}
\big)
\end{equation}
takes the form
\begin{equation}
\begin{aligned}
\bmg^f_t &= \sigma_f \big(W_f [\bmh_{t-\Delta t}, \bmz_t ] + \bmb_f\big) 
&& \bmg^{i}_t = \sigma_i \big( W_i [\bmh_{t-\Delta t}, \bmz_t ] +\bmb_i \big) \\
\tilde{\bmc}_t &=\tanh \big( W_c [\bmh_{t-\Delta t}, \bmz_t ] +\bmb_c \big) 
&& \bmc_t = \bmg^f_t \odot \bmc_{t-\Delta t} + \bmg^{i}_t \odot \tilde{\bmc}_t   \\
\bmg^{\bmz}_t &= \sigma_h \big( W_h [\bmh_{t-\Delta t}, \bmz_t ] + \bmb_h \big) 
&& \bmh_t =  \bmg^{\bmz}_t \odot  \tanh(\bmc_t),
\end{aligned}
\label{eq:lstmequations}
\end{equation}
where $\bmg^f_t, \bmg^{i}_t, \bmg^{\bmz}_t \in \mathbb{R}^{d_{\bmh}}$,
are the gate vector signals (forget, input and output gates),
$\bmz_{t} \in \mathbb{R}^{d_{\bmz}}$ is the latent input at time $t$,
$\bmh_{t} \in \mathbb{R}^{d_{\bmh}}$ is the hidden state,
$\bmc_{t}\in \mathbb{R}^{d_{\bmh}}$ is the cell state,
while $W_f$, $W_i$, $W_c, W_h$ $\in \mathbb{R}^{d_{\bmh} \times (d_{\bmh}+d_{\bmz})}$,
are weight matrices and $\bmb_f, \bmb_i, \bmb_c, \bmb_h \in \mathbb{R}^{d_{\bmh}}$  biases.
The symbol $\odot$ denotes the element-wise product.
The activation functions $\sigma_f$, $\sigma_i$ and $\sigma_h$ are sigmoids.
The dimension of the hidden state $d_{\bmh}$ (number of hidden units) controls the capacity of the cell to encode history information.
The set of trainable parameters of the recurrent mapping $\calH^{
\bmw_{\calH}}$ is thus given by
\begin{equation}
\calH^{
\bmw_{\calH}}
= 
\{
\bmb_f, \bmb_i, \bmb_c, \bmb_h,
W_f, W_i, W_c, W_h
\}
\end{equation}

\clearpage
\newpage

\section{Comparison Measures}
\label{sec:measures}

In this section, we elaborate on the metrics used to quantify the effectiveness of the proposed approach to capture the dynamics and the state statistics of the systems under study.
The mean normalized absolute difference (MNAD) is used to quantify the prediction performance of a method in a deterministic system.
This metric was selected to facilitate comparison of LED with equation-free variants~\cite{lee2020coarse}.
The Wasserstein distance (WD) and the L1-Norm histogram distance (L1-NHD) are utilized to quantify the difference between distributions.
These metrics are used in stochastic systems or in the comparison of state distributions.


\subsection{Mean normalised absolute difference (MNAD)}
\label{sec:mnad}

Assume that a model is used to predict a spatiotemporal field $y(x,t)$, at discrete state $x_i$ and time $t_j$ locations.
Predicted values from a model (neural network, etc.) are denoted with $\tilde{y}$, while the groundtruth (simulation of the equations with a solver based on first principles) with $y$.
The normalized absolute difference (NAD) between the model output and the groundtruth is defined as
\begin{equation}
\text{NAD}(t_j) = \frac{1}{N_x} \sum_{i=1}^{N_x} \frac{ |y(x_i,t_j) - \hat{y}(x_i, t_j) |}{ \operatorname{max}_{i, j}(y(x_i, t_j)) - \operatorname{min}_{i, j}(y(x_i, t_j))},
\label{eq:nad}
\end{equation}
where $N_x$ is the dimensionality of the discretized state $x$.
The NAD depends on the time $t_j$.
The mean NAD (MNAD) is given by the mean over time of the NAD score, i.e.
\begin{equation}
\text{MNAD} = \frac{1}{N_T} \sum_{j=1}^{N_T} \text{NAD}(t_j),
\label{eq:mnad}
\end{equation}
where $N_T$ is the number of time-steps considered.
The MNAD is used in the FitzHugh-Nagumo equation, and the Kuramoto-Sivashinsky equation, to quantify the prediction accuracy of LED and benchmark against other methods (e.g. other propagators on the latent space) or against other equation-free variants.

\subsection{Pearson Correlation Coefficient}
\label{sec:correlation}

Assume as before, the spatiotemporal field $y(x,t)$, at discrete state $x_i$ and time $t_j$ locations.
This can be vectorized in $y_{vec}=\operatorname{vec}(y(x,t)) \in \mathbb{R}^{N_x \cdot N_t \times 1}$.
The same applies to the vectorized prediction $\tilde{y}_{vec}=\operatorname{vec}( \tilde{y}(x,t)) \in \mathbb{R}^{N_x \cdot N_t \times 1}$.
We can compute the Pearson correlation coefficient, or simply correlation, as
\begin{equation}
\text{Correlation} = \frac{\operatorname{COV} \big(y_{vec}, \tilde{y}_{vec}  \big) }{
\sigma_{y_{vec}}
\sigma_{\tilde{y}_{vec}}
},
\label{eq:correlation}
\end{equation}
where $\operatorname{COV}$ is the covariance, and $\sigma$ is the standard deviation.

The correlation is used as a prediction performance metric in the Kuramoto-Sivashinsky equation.

\subsection{Wasserstein Distance}
\label{sec:wasserstein}

The Wasserstein distance (WD), is a metric used to quantify the difference between the distribution functions of two random variables.
It is defined as the integral of the absolute difference of the inverse Cumulative Distribution Functions (CDF) of the random variables.
Assuming two random variables $Z_1$ and $Z_2$, with CDFs given by $\tau = F_{Z_1}(z)$ and $F_{Z_2}(z)$, with $\tau \in [0,1]$, the  Wasserstein metric is defined as
\begin{equation}
\operatorname{WD}(Z_1, Z_2) = \int_{0}^{1} | F_{Z_1}^{-1}(\tau)  -  F_{Z_2}^{-1}(\tau)| \, d \tau.
\end{equation}
In high-dimensional problems, where the random variable is multivariate (random vector), we are reporting the mean WD of each variable after marginalization of all others.

\subsection{L1-Norm Histogram Distance}
\label{sec:l1histogram}

In order to quantify the difference of the distributions of two random multivariate random variables $Z_1$ and $Z_2$, we employ in addition to the WD, the L1-Norm histogram distance.
We measure this metric based on the L1 norm of the difference between the normalized histograms of the random variables computed on the same grid.
The number of bins for the computation of the histograms, is selected according to Rice rule, given by $ N_{bins} = 
\ceil[\big]{\, 2 \sqrt[3]{n} \,}$ where $n$ is the number of observations in the sample $z$.
The WD and the L1-NHD are used to measure the difference between the spatial particle distributions in the Advection-Diffusion model.

\clearpage
\newpage


\section{Results}
\label{sec:results}



\begin{figure*}[tbhp]
\centering
\includegraphics[width=0.9\textwidth]{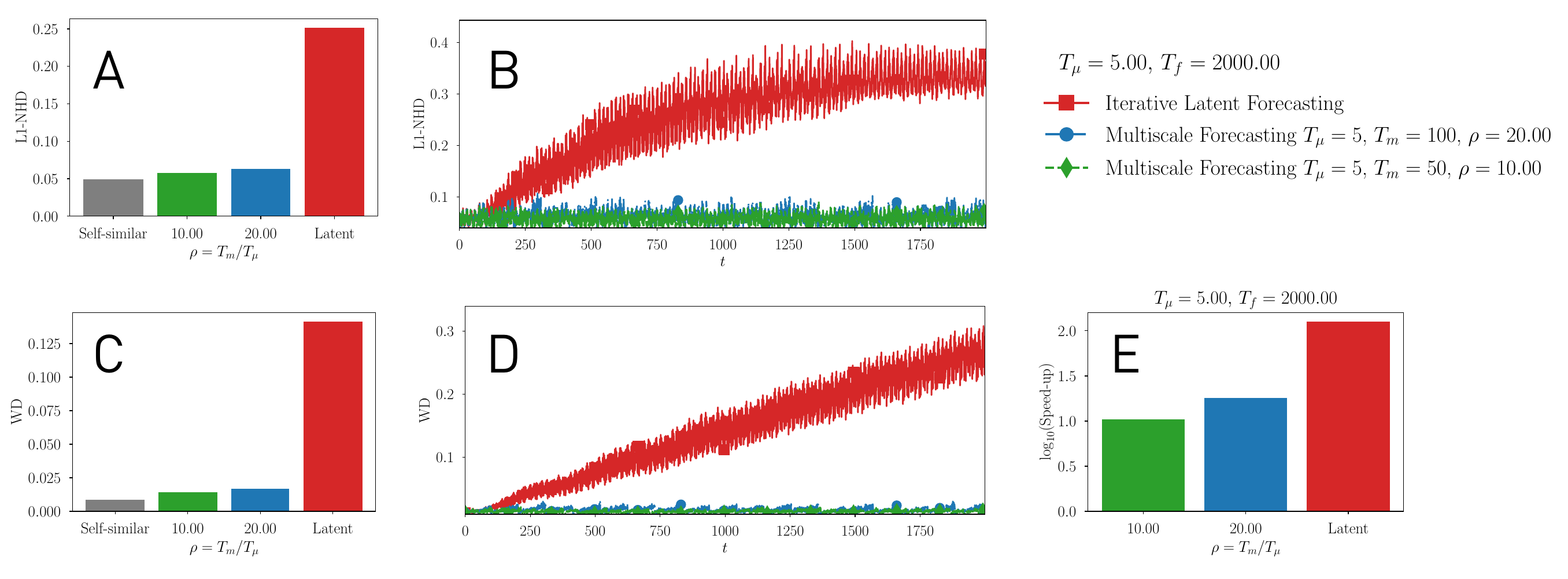}
\caption{
\textbf{A)} The L1-Norm Histogram distance averaged over time and initial conditions.
The self-similar error is plotted for reference, as errors below this level are statistically insignificant.
\textbf{B)} The evolution of the L1-Norm Histogram distance in time averaged over initial conditions.
\textbf{C)} The Wasserstein distance averaged over time and initial conditions.
\textbf{D)} The evolution of the Wasserstein distance in time averaged over initial conditions.
\textbf{E)} The speed-up of LED compared to the micro scale solver is plotted w.r.t. $\rho$.
}
\label{fig:AD3D_figure_keynote_analysis}
\end{figure*}

\subsection{LED for Advection-Diffusion Equation}
\label{app:sec:dif3d}

The LED method is applied to the simulation of the advection-diffusion equation.
The microscale description of the Advection-Diffusion (AD) process is modeled with a system of $N=1000$ particles on a bounded domain $\Omega = \left[-L/2, L/2\right]^{d_{\bm{x}}}$. The particle dynamics are modeled with the stochastic differential equation (SDE)
\begin{equation}
\mathrm{d} \bm{x}_t = \bm{u}_t \mathrm{d} t + \sqrt{D} \, \mathrm{d} \bm{W}_t,
\label{eq:SDE}\end{equation}
where $\bm{x}_t \in \Omega$ denotes the position of the particle at time $t$, $D\in \mathbb{R}$ is the diffusion coefficient, $\mathrm{d} \bm{W}_t \in \mathbb{R}^{d_{\bm{x}}}$ is a Wiener process, and $\bm{u}_t = \bm{A} \cos( \bm{\omega} t) \in \mathbb{R}^{d_{\bm{x}}}$ is a cosine advection (drift) term.
In the following, the three dimensional space $d_{\bm{x}}=3$ is considered, with $D=0.1$, $\bm{A}=[1, 1.74, 0.0]^T$, $\bm{\omega}=[0.2, 1.0, 0.5]^T$, and a domain size of $L=1$.
The P\'eclet number quantifies the rate of advection by the rate of diffusion, i.e. $Pe= \frac{L U}{D}$.
In this work, $L=1$, $U=|\bm{A}|_2\approx 2$, suggest a P\'eclet number of $Pe=20$.
\Cref{eq:SDE} is solved with explicit Euler integration with $\Delta t =10^{-2}$, initial conditions $\bm{x}_0 = \bm{0}$, and reflective boundary conditions ensuring that $\bm{x}_t \in \Omega, \forall t$.
The positions of the particles are saved at a coarser time-step $\Delta t = 1$.
Three datasets are generated by starting from randomly selected initial conditions.
The training and validation datasets consist of $500$ samples each, and the test dataset consists of $4000$ samples.
The full state of the system is high-dimensional, i.e. $\bm{s}_t=[\bm{x}^1_t ; \dots ;\bm{x}^N_t ]^T \in \mathbb{R}^{N\times 3}$.

The particles concentrate on a few ``meta-stable'' states, and transition between them, suggesting that the collective dynamics can be captured by a few latent variables.
It is not straightforward, however, to determine a-priori the number of these states and the patterns of collective motion.
LED unravels this information and provides a computationally efficient multiscale model to approximate the system dynamics.
An AE with a permutation invariant input layer with a latent dimension $d_{\bm{z}}$, an MD decoder and a stateful LSTM-RNN are employed to learn and forecast the dynamics on the low-dimensional manifold.


In this case, the latent dimension of LED is tuned to $d_{\bmz}=8$ based on the log-likelihood on the validation data.
Reducing the latent dimension further, caused a decrease in the validation log-likelihood loss, while increasing the latent dimension did not lead to any significant improvement.
Regarding the rest of the LED hyper-parameters, the $\phi$ function consists of $3 \times 50$ layers and $\tanh$ activation, the permutation invariant space has dimension $M=100$ with mean feature function, and the decoder $g$ consists of a network with $3 \times 50$ layers and $\tanh$ activation, reducing the dimensionality to the desired latent state of dimension $d_{\bmz}=8$.  
The decoder is composed of $3\times50$ layers, and a mixture density output layer, with $25$ hidden units, and $5$ kernels outputting the parameters for the mixture coefficients, the means, and the covariance matrices of the $5$ kernels.
The RNN propagating the dynamics in the latent space, is composed of one stateful LSTM layer with $25$ nodes and was trained with BBTT with a sequence length of $100$.

After training the RNN, the efficiency of LED in forecasting the dynamics is tested in $30$ trajectories starting from different initial conditions randomly sampled from the testing data.
The final prediction horizon is set to $T_f=2000$.
The particle spatial distribution predicted by LED is compared against the groundtruth in terms of the L1-Norm Histogram distance (L1-NHD) and the Wasserstein distance (WD).
The results are shown in Figure~\ref{fig:AD3D_figure_keynote_analysis}.
Three LED variants are considered.
The first variant does not evolve the dynamics on the particle level (Latent-LED, $T_m=0$) and its error increases with time and exhibits the highest errors on average.
The second and third variants, (Multiscale-LED), evolve the low order manifold dynamics (coarse scale) for $T_m$ time units, and the particle dynamics (fine scale) for $T_{\mu}=5$ to correct iteratively for the statistical error.
This effect is due to the explicit dependence of the coarse system dynamics in time, as the $\cos(\bm{\omega} t)$ advection term dominates.
Two values for $T_m$ are considered, $T_m=50$ leading to a relative ratio of coarse to fine simulation time of $\rho=T_{m}/T_{\mu}=10$, and another one with $T_m=100$, leading to $\rho=20$.
This incurs additional computational cost induced by the evolution of the high-dimensional state.
The warm-up time is $T_{warm}=100$ for all variants.
As the multiscale ratio $\rho=T_{m}/T_{\mu}$ is increased, spending more time in the latent propagation, the errors gradually increase.
The propagation in the low dimensional latent space is far less computationally expensive compared to the evolution of the high-dimensional dynamics.
As $\rho$ is increased, greater computational savings are achieved, albeit at the cost of higher approximation error, as depicted in Figure~\ref{fig:AD3D_figure_keynote_analysis}.
The LED is able to generalize to different numbers of particles, as demonstrated in the Section \ref{app:sec:ad3d:generalization}.

\begin{figure*}[ht]
\centering
\includegraphics[width=0.75\textwidth]{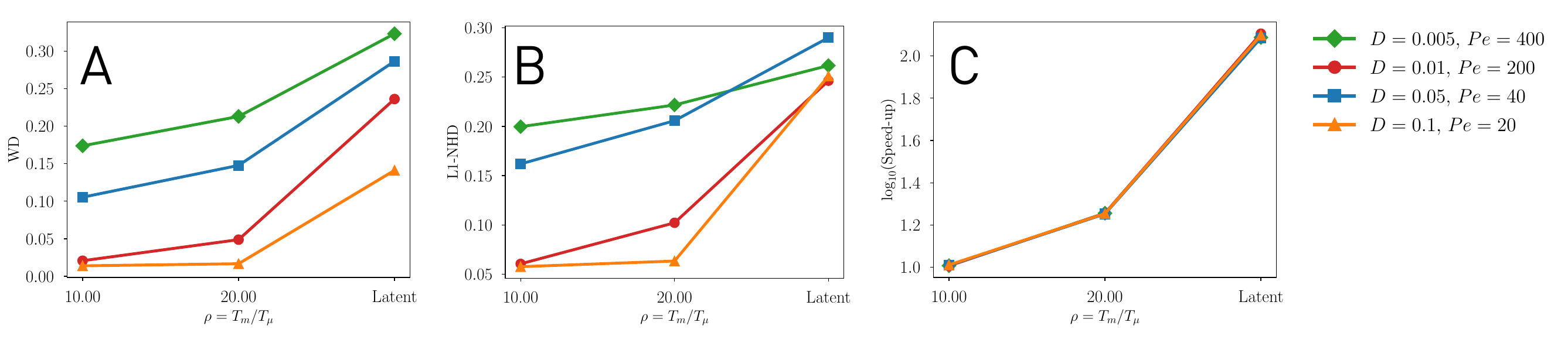}
\caption{
Analysis of the performance of LED for different P\'eclet numbers $Pe \in \{20,40,200,400\}$.
Three LED variants are considered, Latent-LED ($T_m= 0$), and two variants of Multilscale-LED with $T_{\mu}=5$ and $T_m \in \{ 50, 100 \}$.
The warm-up time is $T_{warm}= 100$ for all variants.
\textbf{A)} The Wasserstein distance and \textbf{B)} L1-Norm Histogram distance between the particle spatial distributions averaged over time and initial conditions, plotted with respect to the multiscale ratio $\rho$.
The methods consistently exhibit lower error as the P\'eclet number decreases.
\textbf{C)} The speed-up is plotted w.r.t. $\rho$.
}
\label{fig:AD3D_figure_keynote_analysis_diffusion_coeff}
\end{figure*}

The effectiveness of LED depending on the diffusion coefficient $D$ is shown in Figure~\ref{fig:AD3D_figure_keynote_analysis_diffusion_coeff}.
LED exhibits consistently lower error as the P\'eclet number decreases.
In lower P\'eclet numbers, diffusion becomes dominant.
Since the diffusion is isotropic, it brings the system very fast to a mean solution.

\begin{figure*}[tbhp]
\centering
\begin{subfigure}[tbhp]{0.9\textwidth}
\centering
\includegraphics[width=0.9\textwidth, center]{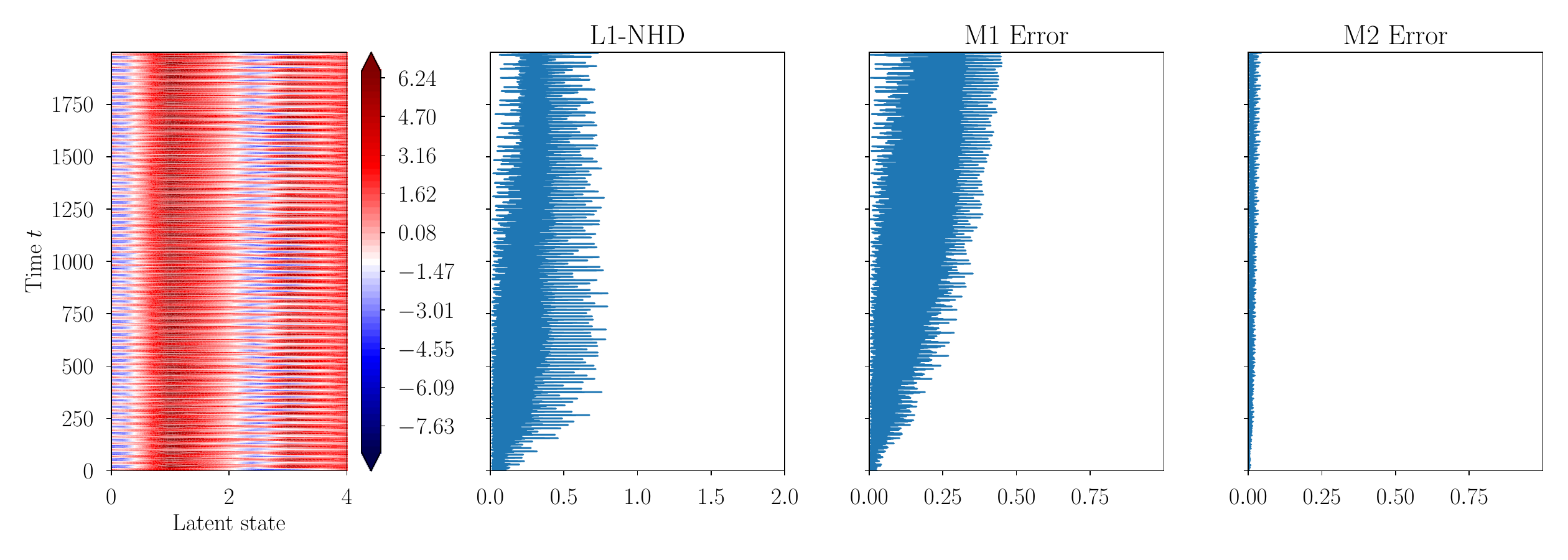}
\caption{LED $T_{\mu}=0$}
\label{fig:app:AD-3D:AD-3D-iterative_latent_forecasting_test_1_density_contour}
\end{subfigure}
\hfill
\centering
\begin{subfigure}[tbhp]{0.9\textwidth}
\centering
\includegraphics[width=1.0\textwidth, center]{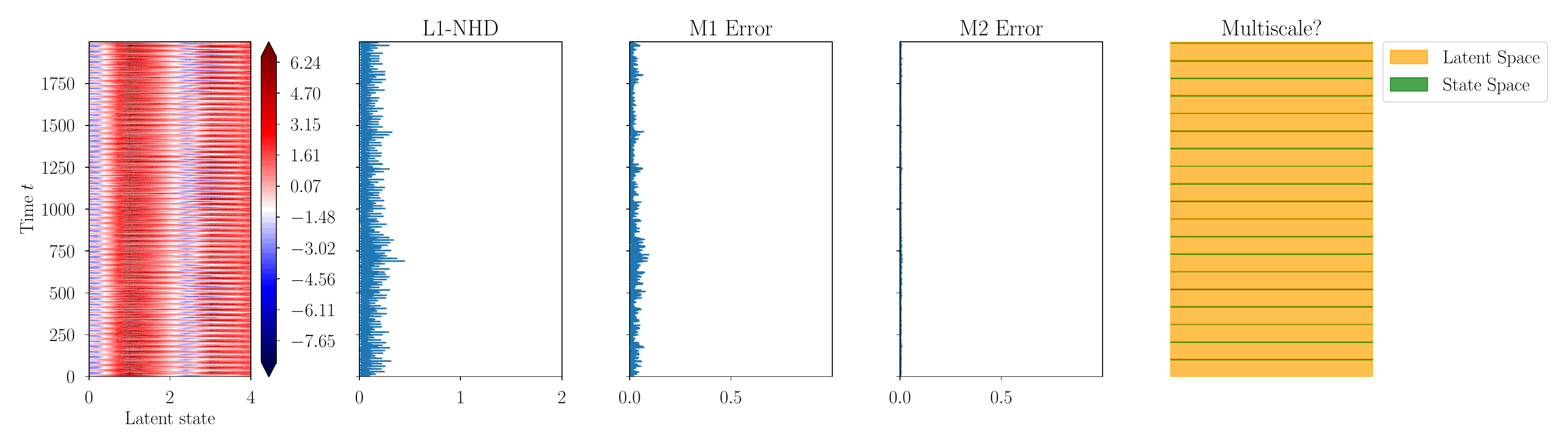}
\caption{LED $T_{m}=100$, $T_{\mu}=5$, $\rho=20$}
\label{fig:app:AD-3D:AD-3D-multiscale_forecasting_micro_5_macro_100_test_1_density_contour}
\end{subfigure}
\centering
\caption{
\textbf{(a)} LED applied on the 3-dimensional Advection-Diffusion equation, iteratively forecasting the evolution of the particles starting from an initial condition in the test data.
The initial warm-up period of LED is set to $T_{warm}=100$.
An AE with a permutation invariant input layer, and a latent dimension of $d_{\bmz}=8$ is utilized to coarse-grain the high-dimensional dynamics.
The decoder of LED is mapping from the latent space to the particle configuration using a MD decoder.
We plot the evolution of the latent state in time, along with the L1-NHD between the predicted and groundtruth particle distributions and the absolute error on the mean, and the standard deviation of the particle distributions. 
LED can forecast the evolution of the particle positions with low error, even though the total dimensionality of the original state describing the configuration of the $N=1000$ particles of the system is $\bm{s}_t \in \mathbb{R}^{1000 \times 3}$.
The network, learned an $d_{\bmz}=8$ dimensional coarse-grained representation of this configuration.
However, due to the iterative prediction with LED, the error on the predicted distribution of particles is increasing with time.
\textbf{(b)} Multiscale propagation in LED.
To alleviate the iterative error propagation, the multiscale propagation is utilized with $T_{m}=100$, $T_{\mu}=5$, $\rho=20$.
Due to the iterative transition between propagation in the latent space $\bm{z}_t$ of LED for $T_m$ and evolution of the micro-scale particle dynamics for $T_{\mu}$, the effect of iterative statistical error propagation is alleviated.
}
\label{fig:app:AD-3D:LED_forecasting}
\end{figure*}

An example of the evolution of the latent state, the errors on the first two moments, and the  L1-NHD between the groundtruth and the predicted spatial distribution of particles in an iterative prediction on the test data is shown in~\Cref{fig:app:AD-3D:AD-3D-iterative_latent_forecasting_test_1_density_contour}.
The initial warm-up period of LED is set to $T_{warm}=100$.
LED captures the variance of the particle positions but due to the iterative error propagation the error on the distribution (L1-NHD and mean position) is increasing with time.


\begin{figure*}[ht]
\centering
\includegraphics[width=0.8\textwidth]{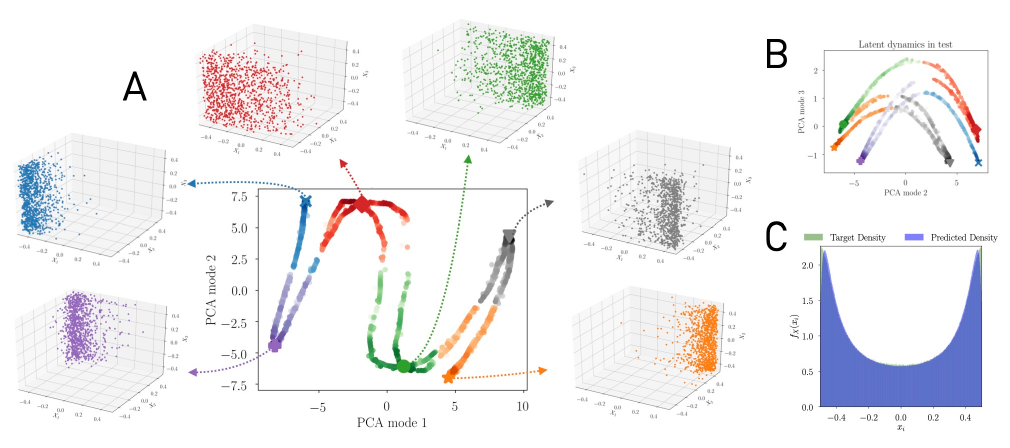}
\caption{
\textbf{A)} Evolution of the second PCA mode of the latent state $\bmz_t \in \mathbb{R}^{d_{\bmz}=8}$, against the first mode.
Higher color intensity denotes higher density.
Six high density regions are identified.
Spectral clustering on the PCA modes of the latent dynamics reveals the clusters.
The six cluster centers are marked, while color illustrates the cluster membership.
The LED probabilistic decoder is employed to map each cluster center to a realization of a high-dimensional simulation state.
LED effectively unravels six meta stable states of the Advection-Diffusion equation, along with the transitions between them, representing the low order effective dynamics.
\textbf{B)} Evolution of the third PCA mode against the second one, colored according to cluster assignment.
\textbf{C)} Density of the particle positions from simulation plotted against the distribution of the positions predicted by LED.
We remark the good agreement between the two distributions.
}
\label{fig:AD3D_figure_keynote_latent_space}
\end{figure*}

In~\Cref{fig:AD3D_figure_keynote_latent_space}, the latent space of LED is clustered to identify frequently visited metastable states that can be mapped back to their respective particle configurations using the decoder.

\clearpage
\newpage

\subsubsection{Hyper-parameter Tuning}
\label{app:sec:ad3d:hyperparams}

The hyper-parameters of LED are given in~\Cref{app:tab:hypauto_ad3d} for the Autoencoder, and~\Cref{app:tab:hyprnn_ad3d} for the RNN.
\begin{table}[tbhp]
\caption{Autoencoder hyper-parameters for Advection-Diffusion in 3-D ($d_{\bmx}=3$)}
\label{app:tab:hypauto_ad3d}
\centering
\begin{tabular}{ |c|c|c| } 
\hline
\text{Hyper-parameter} & 
\text{Values} \\  \hline \hline
Number of AE layers & $\{3\}$ \\
Size of AE layers & $\{ 50 \}$ \\
Activation of AE layers & $\operatorname{tanh}(\cdot)$ \\
Latent dimension & $\{1,2,3,4,5,6,7,8,9,10,12,16,18,22,24,28,32,64 \}$ \\
Residual connections & False \\
Variational & True/False \\
Permutation Invariant Layer $d_p$& $\{ 200, 1001\}$ \\
Number of MD kernels $K$& $\{ 5\}$ \\
Hidden units of MD decoder & $\{ 50\}$ \\
Input/Output data scaling & Min-Max in $[0,1]$ \\
Noise level in the data & $\{0, 1,  10 \}$ $(\permil)$ \\
Weight decay rate & $\{0.0, 0.00001 \}$ \\
Batch size& $32$ \\
Initial learning rate & $0.001$ \\
\hline
\end{tabular}
\end{table}
\begin{table}[tbhp]
\caption{LED-RNN hyper-parameters for Advection-Diffusion in 3-D ($d_{\bmx}=3$)}
\label{app:tab:hyprnn_ad3d}
\centering
\begin{tabular}{ |c|c|c| } 
\hline
\text{Hyper-parameter} & 
\text{Values} \\  \hline \hline
Number of AE layers & $\{3\}$ \\
Size of AE layers & $\{50 \}$ \\
Activation of AE layers & $\operatorname{tanh}(\cdot)$ \\
Latent dimension & $\{8 \}$ \\
Residual connections & False \\
Variational & False \\
Permutation Invariant Layer $d_p$& $\{ 200\}$ \\
Number of MD kernels $K$& $\{ 5\}$ \\
Hidden units of MD decoder & $\{ 50\}$ \\
Input/Output data scaling & Min-Max in $[0,1]$ \\
Noise level in the data & $\{0\}$ \\
Weight decay rate & $\{0.0 \}$ \\
Batch size& $32$ \\
Initial learning rate & $0.001$ \\
BBTT Sequence length & $\{100 \}$ \\
RNN cell type & $\{ \operatorname{lstm}, \operatorname{gru} \}$ \\
Number of RNN layers & $\{1\}$ \\
Size of RNN layers & $\{25 \}$ \\
Activation of RNN Cell & $\operatorname{tanh}(\cdot)$ \\
\hline
\end{tabular}
\end{table}

\clearpage
\newpage
\subsubsection{Generalization to Different Number of Particles}
\label{app:sec:ad3d:generalization}

\begin{figure*}[tbhp]
\centering
\begin{subfigure}[tbhp]{0.6\textwidth}
\centering
\includegraphics[width=1.0\textwidth]{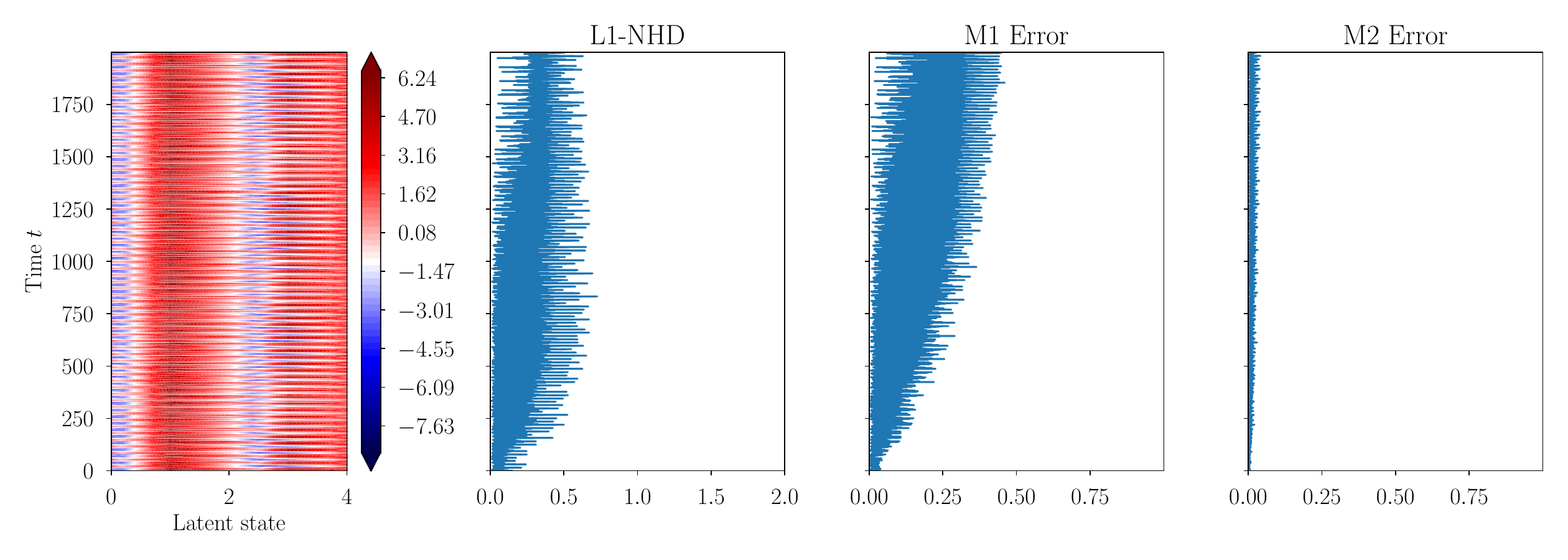}
\caption{LED $T_{\mu=0}$}
\label{fig:app:AD-3D:NPARTICLES_400:AD-3D-N400-iterative_latent_forecasting_test_1_density_contour}
\end{subfigure}
\hfill 
\begin{subfigure}[tbhp]{0.9\textwidth}
\centering
\includegraphics[width=1.0\textwidth]{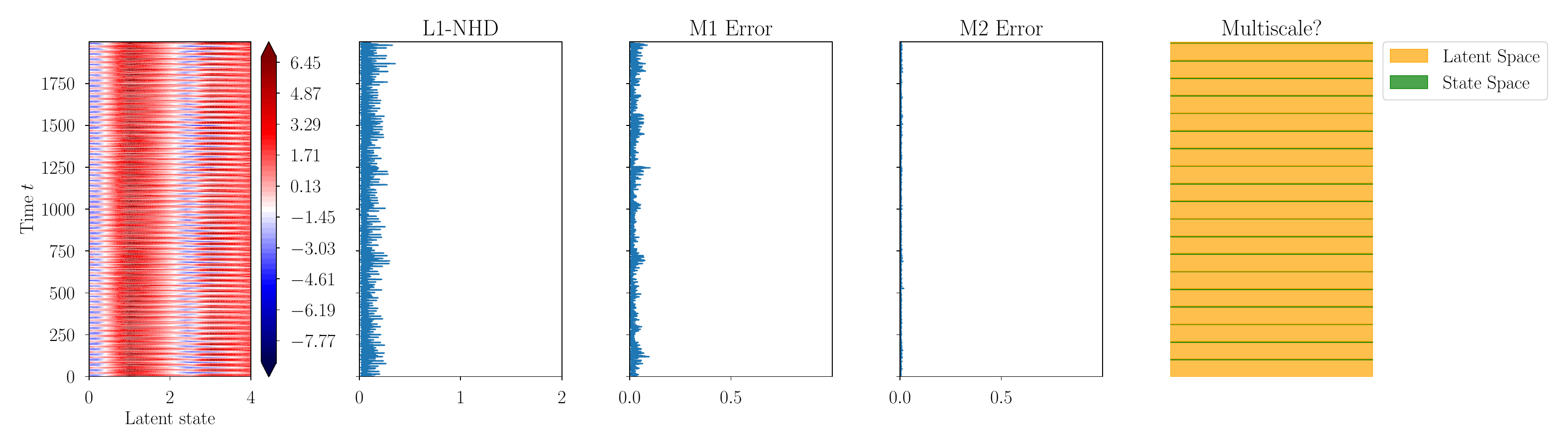}
\caption{LED $T_{\mu}=100, \rho=20$}
\label{fig:app:AD-3D:NPARTICLES_400:AD-3D-N400-multiscale_forecasting_micro_5_macro_100_test_1_density_contour}
\end{subfigure}
\centering
\caption{
LED trained on particle configurations with $N=1000$ number of particles, learned an $d_{\bmz}=8$ dimensional coarse-grained representation of this configuration.
We utilize two models with $T_{\mu}=0$ (iterative latent propagation) and $T_{\mu}=100, \rho=20$ (multiscale forecasting) to forecast the evolution of a particle configuration composed of $N=400$ particles to test the generalization ability of the model.
The initial warm-up period is set to $T_{warm}=100$.
We plot the latent space, the  L1-NHD between the densities of the particle positions, and the error on the first two moments, for both variants of LED.
We observe that the LED is able to successfully generalize to the case of $N=400$ particles.
}
\label{fig:app:AD-3D:NPARTICLES_400:density_contours}
\end{figure*}

In this section, we provide additional results on the \textbf{generalization} of LED for a different \textbf{number of particles} in the simulation.
Due to the permutation invariant encoder, coarse-graining the high-dimensional input of LED, the network is expected to be able to generalize to a different number of particles, since the identified coarse representation should rely on global statistical quantities, and not depend on individual positions.
LED trained in configurations of $N=1000$ particles is utilized to forecast the evolution of $N=400$ particles evolving according to the Advection-Diffusion equation.
The propagation of the errors is plotted in~\Cref{fig:app:AD-3D:NPARTICLES_400:density_contours}.
The initial warm-up period of LED is set to $T_{warm}=100$ for all variants.
We observe an excellent generalization ability of the network.

\clearpage
\newpage



\subsection{FitzHugh-Nagumo Model (FHN)}
\label{app:sec:fhnmodel}

The hyper-parameters of the Autoencoder are reported in~\Cref{app:sec:fhn_hp}.
Input and output are scaled to $[0,1]$ and an output activation function of the form $1+0.5 \tanh(\cdot)$ is used to ensure that the data at the output lie at this range.
The architecture of the CNN we employed is given in~\Cref{fig:FHN_CNN}.
In this case, the inhibitor and activator density are considered two channels of the CNN.

\begin{figure}[tbhp]
\centering
\includegraphics[width=0.9\textwidth]{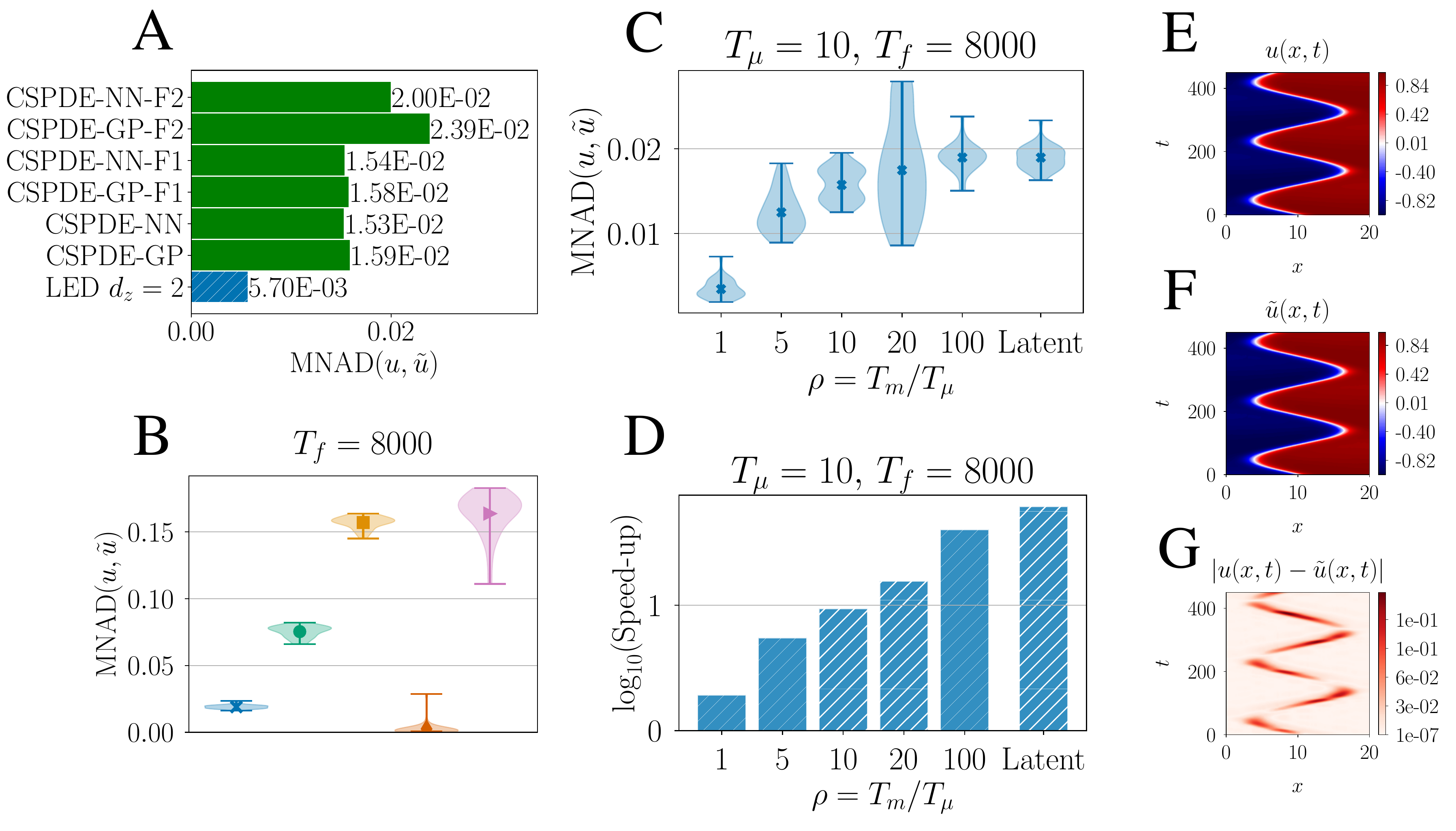}
\caption{
\textbf{A)} Comparison of Latent-LED with $d_{\bm{z}}=2$ with equation-free variants from~\cite{lee2020coarse} in forecasting the dynamics of FHN starting from one initial condition from the testing dataset.
The density and mean MNAD error (averaged over $32$ initial conditions) between the predicted and ground-truth evolution of the activator density is plotted.
\textbf{B)} Comparison of different macrodynamics propagators (\addlegend{L_cross_blue.pdf}{0.95} AE-LSTMend2end; \addlegend{L_circle_tirquaz.pdf}{0.95} AE-LSTM; \addlegend{L_rectange_yellow.pdf}{0.95} AE-MLP; \addlegend{L_diamond_orange.pdf}{0.95} AE-RC; \addlegend{L_tiltedtriangle_purple.pdf}{0.95} AE-SINDy) in iterative latent forecasting.
The MNAD error on the activator density is plotted.
\textbf{C)} The activator mean MNAD error (averaged over $32$ initial conditions) in multiscale forecasting with LED (AE-LSTMend2end with $d_{\bm{z}}=2$)  and its density plotted as a function of the macro-to-micro ratio $\rho=T_m/T_{\mu}$.
\textbf{D)} The speed-up of LED (AE-LSTMend2end with $d_{\bm{z}}=2$) compared to the LB solver plotted w.r.t. $\rho$.
The results for Latent-LED ($T_{\mu}=0$) are denoted with the label ``Latent''.
As $T_m$ is increased (increase $\rho$), the speed-up is increased, albeit at the cost of an increasing MNAD error.
\textbf{E)} The evolution of the activator density in time starting from an initial condition in the test data, along with \textbf{F)} the prediction of Latent-LED and \textbf{G)} absolute difference.
}
\label{fig:FHN:FHN_act}
\end{figure}

\clearpage

\subsubsection{Hyper-parameters and Training Time}
\label{app:sec:fhn_hp}
\begin{figure}[tbhp]
\centering
\includegraphics[width=0.9\textwidth]{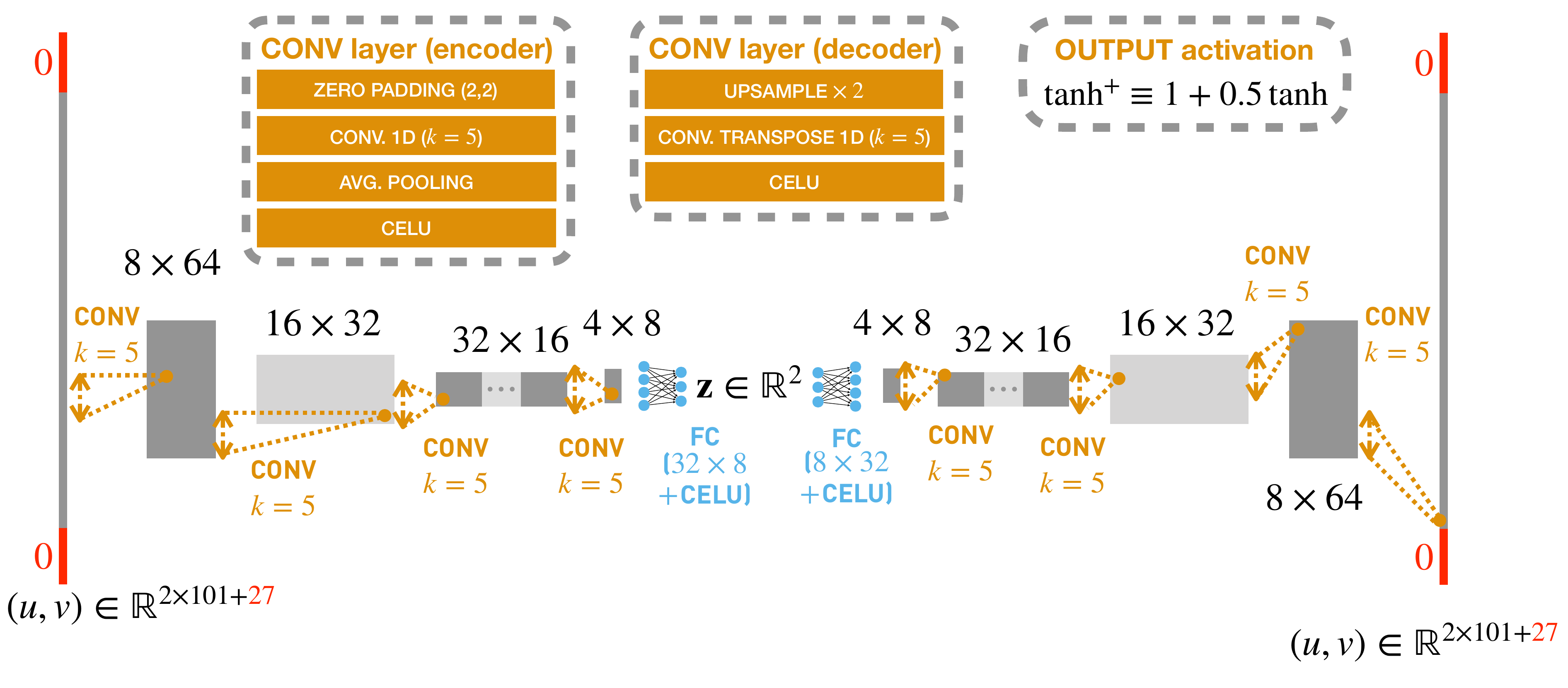}
\caption{
The architecture of the CNN employed in the FHN equation.
First, the input is padded to the closest power of two.
Then, four layers of consecutive application of 1D convolutions, average pooling, CELU activations functions and dropout are used.
Then an MLP is utilized to project to the low-order latent space.
The output activation of the MLP is also $1+0.5 \tanh(\cdot)$.
}
\label{fig:FHN_CNN}
\end{figure}

The hyper-parameter tuning of the autoencoder of LED and training times are reported in~\Cref{app:tab:hypauto_fhn}.
PCA and Diffusion maps have very short fitting (training) times of approximately one minute.
The layers of the CNN autoencoder employed in the FHN and its training times are given in~\Cref{app:tab:fhn:cnn}.
The architecture of the CNN autoencoder employed in the FHN is depicted in~\Cref{fig:FHN_CNN}.

The hyper-parameters for the LSTM and its training times are given in~\Cref{app:tab:hyprnn_fhn}.
For the MLP, a three layered network with CELU activations is employed.
Training time for the MLP is 100 minutes.
The hyper-parameters and training times for the RC are given in~\Cref{app:tab:hyprc_fhn}.
The hyper-parameters and training times for SINDy are given in~\Cref{app:tab:hypsindy_fhn}.

In all cases, the parameters of the best performing model on the validation data is denoted with red color.
\begin{table}[tbhp]
\caption{Autoencoder hyper-parameters and training times for FHN}
\label{app:tab:hypauto_fhn}
\centering
\begin{tabular}{ |c|c|c| } 
\hline
\text{Hyper-parameter tuning} & 
\text{Values} \\  \hline \hline
Number of AE layers & $\{3 \}$ \\
Size of AE layers & $\{100 \}$ \\
Activation of AE layers & $\operatorname{celu}(\cdot)$ \\
Latent dimension & $\{1, {\color{red}2}, 3, 4, 5, 6, 7, 8, 9, 10, 11, 12, 16, 20, 24, 28, 32, 36, 40, 64\}$ \\
Input/Output data scaling & $[0,1]$ \\
Output activation & $1+0.5 \tanh(\cdot)$ \\
Weight decay rate & $\{ {\color{red}0.0}, 0.0001\}$ \\
Batch size& $32$ \\
Initial learning rate & $0.001$ \\
\hline
\end{tabular}
\\[0.4cm]
\begin{tabular}{ |c|c|c| } 
\multicolumn{3}{c}{Training times [minutes]} \\ \hline
\text{Min} & \text{Mean} & \text{Max} \\  \hline \hline
130 & 180 & 235 \\
\hline
\end{tabular}
\end{table}

\begin{table}[tbhp]
\caption{LED-RNN hyper-parameters and training times for FHN}
\label{app:tab:hyprnn_fhn}
\centering
\begin{tabular}{ |c|c|c| } 
\hline
\text{Hyper-parameter} & 
\text{Values} \\  \hline \hline
end2end training & {\color{red} True}  / False \\
Number of AE layers & $\{3\}$ \\
Size of AE layers & $\{100 \}$ \\
Activation of AE layers & $\operatorname{celu}(\cdot)$ \\
Latent dimension & $ 2$ \\
Input/Output data scaling & $[0,1]$ \\
Output activation & $1+0.5 \tanh(\cdot)$ \\
Weight decay rate & $ 0.0$ \\
Batch size& $32$ \\
Initial learning rate & $0.001$ \\
BPTT Sequence length & $\{20,  {\color{red}40}, 60 \}$ \\
Output forecasting loss & True/False \\
RNN cell type & $\operatorname{lstm}$ \\
Number of RNN layers & $1$ \\
Size of RNN layers & $\{16, {\color{red}32}, 64 \}$ \\
Activation of RNN Cell & $\operatorname{tanh}(\cdot)$ \\
Output activation of RNN Cell & $1+0.5 \tanh(\cdot)$ \\
\hline
\end{tabular}
\\[0.4cm]
\begin{tabular}{ |c||c|c|c| } \hline
Training times [minutes]&\text{Min} & \text{Mean} & \text{Max} \\  \hline \hline
end2end training &2.2 & 2.5 & 2.8 \\
only the RNN (sequential) &0.9 & 1.2 & 1.6 \\
\hline
\end{tabular}
\end{table}

\begin{table}[tbhp]
\caption{CNN Autoencoder and training times for FHN}
\label{app:tab:fhn:cnn}
\centering
\begin{tabular}{ |c|c| } 
\hline
Layer & \text{ENCODER}  \\ \hline
  (0) & ConstantPad1d(padding=(13, 14), value=0.0)  \\
  (1) & ConstantPad1d(padding=(2, 2), value=0.0)  \\
  (2) & Conv1d(2, 8, kernel\_size=(5,), stride=(1,))  \\
  (3) & AvgPool1d(kernel\_size=(2,), stride=(2,), padding=(0,))  \\
  (4) & CELU(alpha=1.0)  \\
  (5) & ConstantPad1d(padding=(2, 2), value=0.0)  \\
  (6) & Conv1d(8, 16, kernel\_size=(5,), stride=(1,))  \\
  (7) & AvgPool1d(kernel\_size=(2,), stride=(2,), padding=(0,))  \\
  (8) & CELU(alpha=1.0)  \\
  (9) & ConstantPad1d(padding=(2, 2), value=0.0)  \\
  (10) & Conv1d(16, 32, kernel\_size=(5,), stride=(1,))  \\
  (11) & AvgPool1d(kernel\_size=(2,), stride=(2,), padding=(0,))  \\
  (12) & CELU(alpha=1.0)  \\
  (13) & ConstantPad1d(padding=(2, 2), value=0.0)  \\
  (14) & Conv1d(32, 4, kernel\_size=(5,), stride=(1,))  \\
  (15) & AvgPool1d(kernel\_size=(2,), stride=(2,), padding=(0,))  \\
  (16) & Flatten(start\_dim=-2, end\_dim=-1)  \\
  (17) & Linear(in\_features=32, out\_features=$d_{\mathbf{z}}$, bias=True)  \\
  (18) & CELU(alpha=1.0)  \\
  & $\mathbf{z} \in \mathbb{R}^{d_{\mathbf{z}}}$ \\ \hline  \hline
Layer & \text{DECODER}  \\ \hline 
  (1) & Linear(in\_features=$d_{\mathbf{z}}$, out\_features=32, bias=True) \\
  (2) & CELU(alpha=1.0)  \\
  (3) & Upsample(scale\_factor=2.0, mode=linear) \\
  (4) & ConvTranspose1d(4, 32, kernel\_size=(5,), stride=(1,), padding=(2,)) \\
  (5) & CELU(alpha=1.0) \\
  (6) & Upsample(scale\_factor=2.0, mode=linear) \\
  (7) & ConvTranspose1d(32, 16, kernel\_size=(5,), stride=(1,), padding=(2,)) \\
  (8) & CELU(alpha=1.0) \\
  (9) & Upsample(scale\_factor=2.0, mode=linear) \\
  (10) & ConvTranspose1d(16, 8, kernel\_size=(5,), stride=(1,), padding=(2,)) \\
  (11) & CELU(alpha=1.0) \\
  (12) & Upsample(scale\_factor=2.0, mode=linear) \\
  (13) & ConvTranspose1d(8, 2, kernel\_size=(5,), stride=(1,), padding=(2,)) \\
 (14) & 1 + 0.5 Tanh() \\
  (15) & Unpad() \\
\hline  \hline 
Latent dimension $d_{\mathbf{z}}$  & $\{1,2, 3, 4, 5, 6, 7, 8, 9, 10, 11, 12, 16, 20, 24, 28, 32, 36, 40, 64\}$ \\ \hline
\end{tabular}
\\[0.4cm]
\begin{tabular}{ |c|c|c| } 
\multicolumn{3}{c}{Training times [minutes]} \\ \hline
\text{Min} & \text{Mean} & \text{Max} \\  \hline \hline
215 & 370 & 530 \\
\hline
\end{tabular}
\end{table}


\begin{table}[tbhp]
\caption{Reservoir Computer hyper-parameters and training times (in CNN-RC) for FHN}
\label{app:tab:hyprc_fhn}
\centering
\begin{tabular}{ |c|c|c| } 
\hline
\text{Hyper-parameter tuning} & 
\text{Values} \\  \hline \hline
Solver & Pseudoinverse \\
Size & $1000$ \\
Degree & $10$ \\
Radius & $0.99$ \\
Input scaling $\sigma$ & $\{0.5, {\color{red} 1}, 2 \}$ \\
Dynamics length & $100$ \\
Regularization $\eta$ & $\{0.0, 0.001, 0.0001, {\color{red} 0.00001} \}$ \\
Noise level per mill & $\{ {\color{red} 10}, 20, 30, 40, 100\}$ \\
\hline
\end{tabular}
\\[0.4cm]
\begin{tabular}{ |c|c|c| } 
\multicolumn{3}{c}{Training times [minutes]} \\ \hline
\text{Min} & \text{Mean} & \text{Max} \\  \hline \hline
0.15 & 0.18 & 0.19 \\
\hline
\end{tabular}
\end{table}


\begin{table}[tbhp]
\caption{SINDy hyper-parameters and training times (in CNN-SINDy) for FHN}
\label{app:tab:hypsindy_fhn}
\centering
\begin{tabular}{ |c|c|c| } 
\hline
\text{Hyper-parameter tuning} & 
\text{Values} \\  \hline \hline
Degree & $\{1,2,  {\color{red} 3}\}$ \\
Threshold & $\{0.001, 0.0001, {\color{red} 0.00001} \}$ \\
Library & Polynomials \\
\hline
\end{tabular}
\\[0.4cm]
\begin{tabular}{ |c|c|c| } 
\multicolumn{3}{c}{Training times [minutes]} \\ \hline
\text{Min} & \text{Mean} & \text{Max} \\  \hline \hline
0.14 & 0.23 & 0.32 \\
\hline
\end{tabular}
\end{table}

\clearpage
\newpage



\subsection{Kuramoto-Sivashinsky}
\label{app:sec:kuramoto}

\begin{figure*}[tbhp]
\centering
\includegraphics[width=0.7\textwidth]{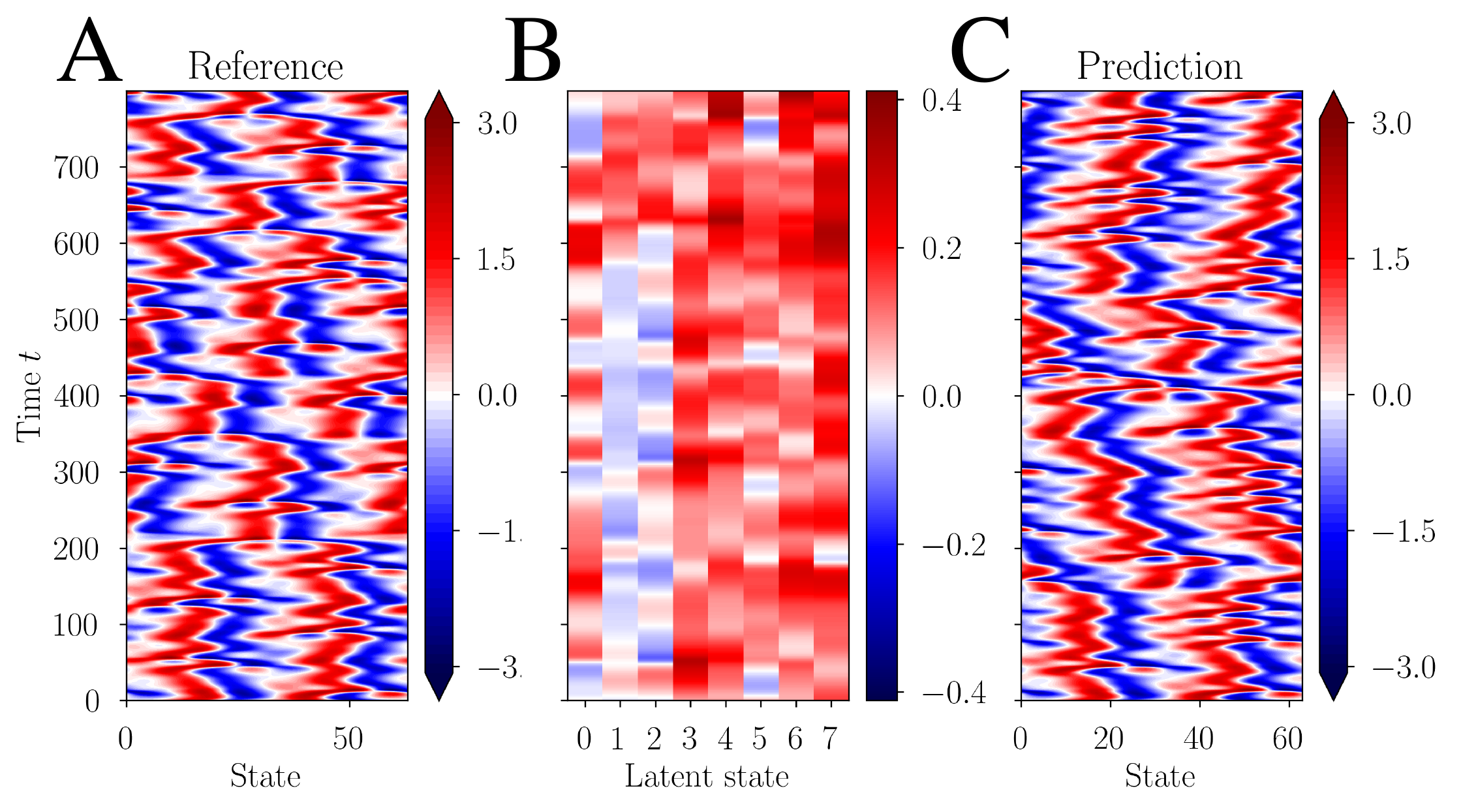}
\caption{
\textbf{A)}
Contour plot of the KS dynamics starting from an initial condition from the test data.
\textbf{B)} The evolution of the $d_{\bm{z}}=8$ dimensional latent state of Latent-LED.
\textbf{C)} The predicted field by Latent-LED iteratively propagating the dynamics on a $d_{\bm{z}}=8$ dimensional latent space, after a warm-up period $T_{warm}=60$ ($T_{\mu}=0$).
}
\label{fig:app:KS:KS_2}
\end{figure*}

A KS trajectory is plotted in~\Cref{fig:app:KS:KS_2}, along with the latent space evolution of Latent-LED and the predicted trajectory.
We observe that the long-term climate is reproduced, although the LED is propagating an $8$-dimensional latent state.

\subsubsection{Hyper-parameters and Training Times}
\label{sec:kshyperparams}

\begin{figure}[tbhp]
\centering
\includegraphics[width=0.9\textwidth]{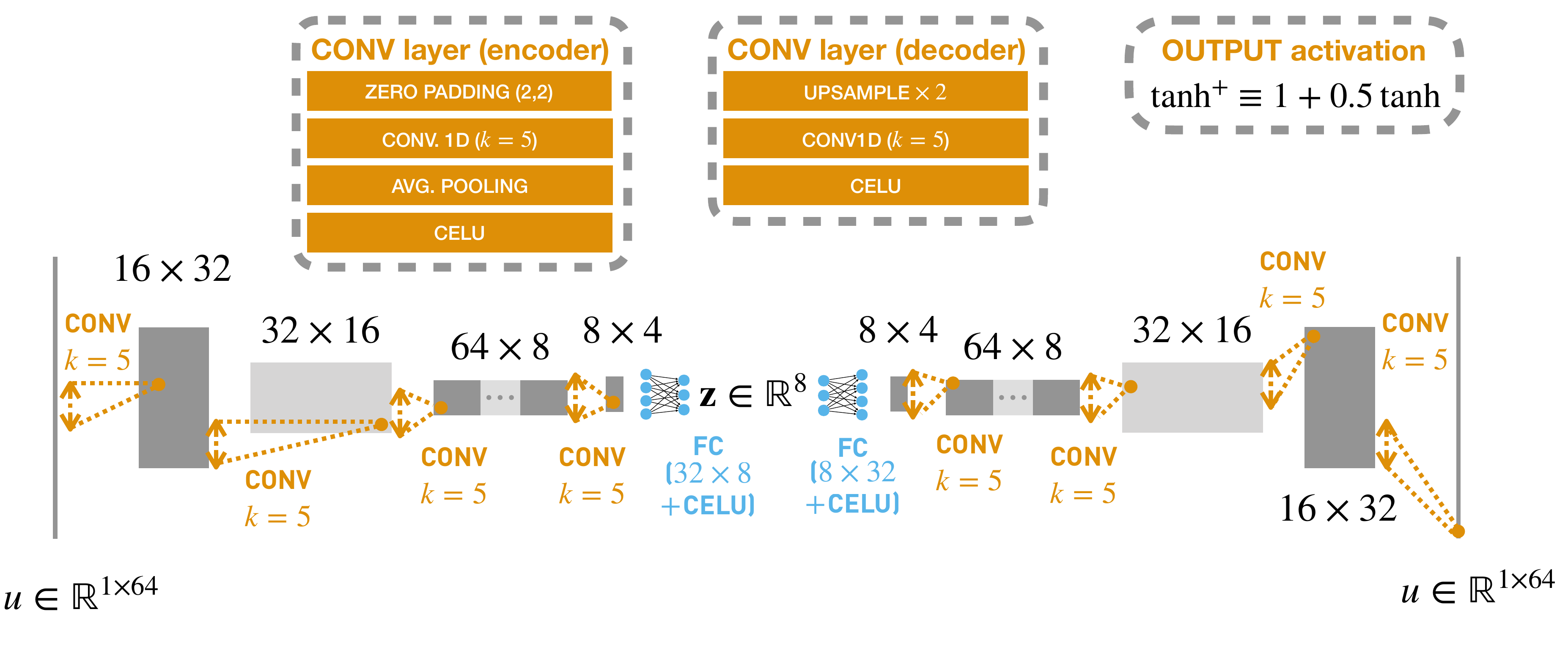}
\caption{
The architecture of the CNN employed in KS.
}
\label{fig:KS_CNN}
\end{figure}

The hyper-parameter tunings and training times for the AE and CNN are given in~\Cref{app:tab:hypauto_ks} and~\Cref{app:tab:hypcnn_ks} respectively .
The architecture of the CNN autoencoder employed in KS is given in~\Cref{app:tab:ks:cnn} along with the training times, and depicted in~\Cref{fig:KS_CNN}.
PCA fitting time is approximately one minute.
The hyper-parameters and training times of the LSTM-RNN of LED are given in~\Cref{app:tab:hyprnn_ks}.
The hyper-parameters and training times of the RC are given in~\Cref{app:tab:hyprc_ks}.
The hyper-parameters and training times of SINDy are given in~\Cref{app:tab:hypsindy_ks}.
\begin{table}[tbhp]
\caption{Autoencoder hyper-parameters for KS}
\label{app:tab:hypauto_ks}
\centering
\begin{tabular}{ |c|c|c| } 
\hline
\text{Hyper-parameter tuning} & 
\text{Values} \\  \hline \hline
Number of AE layers & $\{3 \}$ \\
Size of AE layers & $\{100 \}$ \\
Activation of AE layers & $\operatorname{celu}(\cdot)$ \\
Latent dimension & $\{1, 2, 3, 4, 5, 6, 7, {\color{red} 8}, 9, 10, 11, 12, 16, 20, 24, 28, 32, 36, 40, 64\}$ \\
Input/Output data scaling & $[0,1]$ \\
Output activation & $1+0.5 \tanh(\cdot)$ \\
Weight decay rate & $\{{\color{red} 0.0}, 0.0001\}$ \\
Batch size& $32$ \\
Initial learning rate & $0.001$ \\
\hline
\end{tabular}
\\[0.4cm]
\begin{tabular}{ |c|c|c| } 
\multicolumn{3}{c}{Training times [minutes]} \\ \hline
\text{Min} & \text{Mean} & \text{Max} \\  \hline \hline
160 & 192 & 311 \\
\hline
\end{tabular}
\end{table}

\begin{table}[tbhp]
\caption{CNN hyper-parameters for KS}
\label{app:tab:hypcnn_ks}
\centering
\begin{tabular}{ |c|c|c| } 
\hline
\text{Hyper-parameter} & 
\text{Values} \\  \hline \hline
Convolutional & True \\
Kernels & Encoder: $5- 5-5-5$, Decoder:  $5- 5-5-5$\\
Channels & $1-16-32-64-8-\mathbf{d}_z-8-64-32-16-1$ \\
Batch normalization & True / {\color{red} False} \\
Transpose convolution & True / {\color{red} False} \\
Pooling & Average \\
Activation & $\operatorname{celu}(\cdot)$ \\
Latent dimension & $\{1, 2, 3, 4, 5, 6, 7, {\color{red} 8}, 9, 10, 11, 12, 16, 20, 24, 28, 32, 36, 40, 64\}$ \\
Input/Output data scaling & $[0,1]$ \\
Output activation & $1+0.5 \tanh(\cdot)$ \\
Weight decay rate & $ 0.0$ \\
Batch size& $32$ \\
Initial learning rate & $0.001$ \\
\hline
\end{tabular}
\\[0.4cm]
\begin{tabular}{ |c|c|c| } 
\multicolumn{3}{c}{Training times [minutes]} \\ \hline
\text{Min} & \text{Mean} & \text{Max}  \\  \hline \hline
236 & 311 & 476 \\
\hline
\end{tabular}
\end{table}

\begin{table}[tbhp]
\caption{CNN Autoencoder for KS}
\label{app:tab:ks:cnn}
\centering
\begin{tabular}{ |c|c| } 
\hline
Layer & \text{ENCODER}  \\ \hline
  (1) & ConstantPad1d(padding=(2, 2), value=0.0)  \\
  (2) & Conv1d(1, 16, kernel\_size=(5,), stride=(1,))  \\
  (3) & AvgPool1d(kernel\_size=(2,), stride=(2,), padding=(0,))  \\
  (4) & CELU(alpha=1.0)  \\
  (5) & ConstantPad1d(padding=(2, 2), value=0.0)  \\
  (6) & Conv1d(16, 32, kernel\_size=(5,), stride=(1,))  \\
  (7) & AvgPool1d(kernel\_size=(2,), stride=(2,), padding=(0,))  \\
    (8) & CELU(alpha=1.0)  \\
  (9) & ConstantPad1d(padding=(2, 2), value=0.0)  \\
  (10) & Conv1d(32, 64, kernel\_size=(5,), stride=(1,))  \\
  (11) & AvgPool1d(kernel\_size=(2,), stride=(2,), padding=(0,))  \\
  (12) & CELU(alpha=1.0)  \\
  (13) & ConstantPad1d(padding=(2, 2), value=0.0)  \\
  (14) & Conv1d(64, 8, kernel\_size=(5,), stride=(1,))  \\
  (15) & AvgPool1d(kernel\_size=(2,), stride=(2,), padding=(0,))  \\
    (16) & CELU(alpha=1.0)  \\
  (17) & Flatten( start\_dim =-2, end\_dim = -1)  \\
  (18) & Linear(in\_features =32, out\_features =8, bias=True)  \\
  (19) & CELU(alpha=1.0)  \\
  & $\mathbf{z} \in \mathbb{R}^8$ \\ \hline  \hline
Layer & \text{DECODER}  \\ \hline 
  (1) & Linear(in\_features=8, out\_features=32, bias=True)  \\
  (2) & CELU(alpha=1.0)  \\
  (3) & Upsample(scale\_factor=2.0, mode=linear)  \\
  (4) & Conv1d(8, 64, kernel\_size=(5,), stride=(1,), padding=(2,))  \\
  (5) & CELU(alpha=1.0)  \\
  (6) & Upsample(scale\_factor=2.0, mode=linear)  \\
  (7) & Conv1d(64, 32, kernel\_size=(5,), stride=(1,), padding=(2,))  \\
  (8) & CELU(alpha=1.0)  \\
  (9) & Upsample(scale\_factor=2.0, mode=linear)  \\
  (10) & Conv1d(32, 16, kernel\_size=(5,), stride=(1,), padding=(2,))  \\
  (11) & CELU(alpha=1.0)  \\
  (12) & Upsample(scale\_factor=2.0, mode=linear)  \\
  (13) & Conv1d(16, 1, kernel\_size=(5,), stride=(1,), padding=(2,))  \\
  (14) & 1 + 0.5 Tanh() \\
\hline 
\end{tabular}
\end{table}


\begin{table}[tbhp]
\caption{LED (LSTM-RNN) hyper-parameters and training times for KS}
\label{app:tab:hyprnn_ks}
\centering
\begin{tabular}{ |c|c|c| } 
\hline
\text{Hyper-parameter} & 
\text{Values} \\  \hline \hline
end2end training &{\color{red} False} / True   \\
Convolutional AE (CNN) & True \\
Kernels & Encoder: $5- 5-5-5$, Decoder:  $5- 5-5-5$\\
Channels & $1-16-32-64-8-\mathbf{d}_z-8-64-32-16-1$ \\
Batch normalization & False\\
Transpose convolution & False \\
Pooling & Average \\
Activation & $\operatorname{celu}(\cdot)$ \\
Latent dimension & $8$ \\
Input/Output data scaling & $[0,1]$ \\
Output activation & $1+0.5 \tanh(\cdot)$ \\
Weight decay rate & $ 0.0$ \\
Batch size& $32$ \\
Initial learning rate & $0.001$ \\
BPTT Sequence length & $\{ 25,  {\color{red} 50}, 100 \}$ \\
Output forecasting loss & True/False \\
RNN cell type & $\operatorname{lstm}$ \\
Number of RNN layers & $1$ \\
Size of RNN layers & $\{64, 128, 256, {\color{red} 512} \}$ \\
Activation of RNN Cell & $\operatorname{tanh}(\cdot)$ \\
Output activation of RNN Cell & $1+0.5 \tanh(\cdot)$ \\
\hline
\end{tabular}
\\[0.4cm]
\begin{tabular}{ |c||c|c|c| } \hline
Training times [minutes]&\text{Min} & \text{Mean} & \text{Max}  \\  \hline \hline
end2end training &476& 978 & 1140 \\
only the RNN (sequential) &960 & 1100 & 1140 \\
\hline
\end{tabular}
\end{table}


\begin{table}[tbhp]
\caption{Reservoir Computer hyper-parameters and training times (in CNN-RC) for KS}
\label{app:tab:hyprc_ks}
\centering
\begin{tabular}{ |c|c|c| } 
\hline
\text{Hyper-parameter tuning} & 
\text{Values} \\  \hline \hline
Solver & Pseudoinverse \\
Size & $1000$ \\
Degree & $10$ \\
Radius & $0.99$ \\
Input scaling $\sigma$ & $\{0.5,  1, {\color{red} 2} \}$ \\
Dynamics length & $100$ \\
Regularization $\eta$ & $\{  {\color{red} 0.0}, 0.001, 0.0001,0.00001 \}$ \\
Noise level per mill & $\{  {\color{red} 10}, 20, 30, 40, 100\}$ \\
\hline
\end{tabular}
\\[0.4cm]
\begin{tabular}{ |c|c|c| } 
\multicolumn{3}{c}{Training times [minutes]} \\ \hline
\text{Min} & \text{Mean} & \text{Max}  \\  \hline \hline
0.25 & 0.35 & 0.38 \\
\hline
\end{tabular}
\end{table}

\begin{table}[tbhp]
\caption{SINDy hyper-parameters and training times (in CNN-SINDy) for KS}
\label{app:tab:hypsindy_ks}
\centering
\begin{tabular}{ |c|c|c| } 
\hline
\text{Hyper-parameter tuning} & 
\text{Values} \\  \hline \hline
Library & Polynomials \\
Degree & $\{1,2,  {\color{red} 3}\}$ \\
Threshold & $\{ 0.001, {\color{red} 0.0001}, 0.00001 \}$ \\
\hline
\end{tabular}
\\[0.4cm]
\begin{tabular}{ |c|c|c| } 
\multicolumn{3}{c}{Training times [minutes]} \\ \hline
\text{Min} & \text{Mean} & \text{Max}  \\  \hline \hline
0.13 & 0.62 & 1.59 \\
\hline
\end{tabular}
\end{table}

\clearpage
\newpage



\subsection{Viscous Flow Behind a Cylinder}
\label{app:sec:ns}

The flow behind a cylinder in the two dimensional space is simulated by solving the incompressible Navier-Stokes equations with Brinkman penalization to enforce the no-slip boundary conditions on the surface of the cylinder~\cite{rossinelli2015mrag,Bost2010}, i.e. %
\begin{equation}\label{eq:Incompressible-Navier-Stokes}
\begin{split}
\frac{\partial \mathbf{u}}{\partial t}+(\mathbf{u}\cdot \mathbf{\nabla})\mathbf{u}&=-\frac{\nabla p}{\rho}+\nu\Delta \mathbf{u}+\lambda\chi^{(s)}(\mathbf{u}^{(s)}-\mathbf{u})\,,\\
\mathbf{\nabla}\cdot\mathbf{u}&=0\,,
\end{split}
\end{equation}
where $\mathbf{u} = [ u_x, u_y ]^T \in\mathbb{R}^2$ is the velocity, $p\in\mathbb{R}$ is the pressure field, $\rho$ is the density, $\nu$ is the kinematic viscocity, and $\lambda$ is the penalization coefficient.
The velocity-field $\mathbf{u}^{(s)}\in\mathbb{R}^2$ describes the translation of the cylinder. 
The numerical method of the flow solver is finite differences, with the incompressibility enforced through pressure projection.
The computational domain is $\Omega=[0,1]\times[0,0.5]$, the cylinder is positioned at $(0.2,0.5)\in\Omega$, with diameter $D=0.075$.
The cylinder is described by the characteristic function $\chi^{(s)}$, that is $\chi^{(s)}=1$ inside the cylinder $\Omega^{(s)}$ and $\chi^{(s)}=0$ outside $\Omega\setminus\Omega^{(s)}$.
We consider the application of LED to two Reynolds' numbers $Re=100$ and $Re=1000$, by setting $\nu=0.0001125$ and $\nu=0.00001125$ respectively.
The Strouhal number (defined in the SI Equation 26) is $\operatorname{St}=0.175$, and $\operatorname{St}=0.225$ for $Re=100$ and $Re=1000$ respectively.
For both cases, the domain is discretized using $1024 \times512$ grid-points and the time-step $\delta t$ is adapted to ensure that the CFL number is fixed at $0.5$.
More details on the domain size and simulation are provided in the SI 3 D.

Equation~\ref{eq:Incompressible-Navier-Stokes} is solved for the velocity $\mathbf{u}\in\mathbb{R}^2$ and pressure field $p\in\mathbb{R}$ using the pressure projection method.
First, we perform advection and diffusion of the flow field in the whole domain
\begin{equation}\label{eq:advection-diffusion}
\mathbf{u}^*= \mathbf{u}^t + \delta t \left(\nu \Delta \mathbf{u}^t - (\mathbf{u}^t \cdot \nabla) \mathbf{u}^t \right)\,
.
\end{equation}
The continuity equation requires the field to be divergence-free.
This condition is imposed with the pressure projection
\begin{equation}\label{eq:pressure-projection}
\mathbf{u}^{**}=\mathbf{u}^*-\delta t\frac{\nabla p^{t+1}}{\rho}\,.
\end{equation}
The pressure field used here is obtained by solving the Poisson equation emerging from the divergence of~\Cref{eq:pressure-projection}, i.e.
\begin{equation}
\Delta p^{t+1}=\frac{\rho}{\delta t}\nabla\cdot\mathbf{u}^*\,
.
\end{equation}
Note that adding~\Cref{eq:pressure-projection} and~\Cref{eq:advection-diffusion} yields the original~\Cref{eq:Incompressible-Navier-Stokes} without the penalization term for Euler timestepping.
The time-step is completed by applying the penalization force using $\delta t \lambda=1$,
\begin{equation}\label{eq:penalisation}
\mathbf{u}^{t+1}=\mathbf{u}^{**}+\chi^{(s),t+1}(\mathbf{u}^{(s),t+1}-\mathbf{u}^{**})\,
.
\end{equation}
We remark that the penalisation force acts as a Lagrange multiplier enforcing the translation motion of the cylinder on the fluid.
The temporally discrete equations described above are solved on a grid with spacing $\Delta x$ using second-order central finite differences for diffusion terms, and a third-order upwind scheme for advection terms.

For the simulated impulsively started cylinder the Reynolds-number for a cylinder with diameter $D$ moving with velocity $v$ in a fluid with kinematic viscosity $\nu$ is defined as
\begin{equation}
\label{eq:Re}
Re=\frac{D v}{\nu}.
\end{equation}
In the present simulations the cylinder moves with constant velocity $v=0.15$ in $-x$-direction.
The computational domain is chosen to be $\Omega=[0,1]\times[0,0.5]$ and moves with the center of mass of the sphere with diameter $D=0.075$, that is fixed at $(0.2,0.5)\in\Omega$.
Here we present results for a simulation at $Re=100$ and $Re=1000$ by setting the kinematic viscocity to be $\nu=0.0001125$ and $\nu=0.00001125$ respectively.
For both cases, the domain is discretized using $1024 \times512$ gridpoints and the time-step $\delta t$ is adapted to ensure that the CFL-number is fixed at 0.5.

The Strouhal number $\operatorname{St}$ describes the periodic vortex shedding at the wake of the cylinder.
It is defined as
\begin{equation}
\label{eq:St}
St=\frac{D f}{\nu}.
\end{equation}
where $f$ is the frequency of vortex shedding.
In our case, $\operatorname{St}=0.175$ for $Re=100$, and $\operatorname{St}=0.225$ for $Re=1000$.

The state of the simulation is described by the velocity $\mathbf{u} \in \mathbb{R}^2$ and the pressure $p \in \mathbb{R}$ at each grid point.
The drag coefficient ($Cd$) around the cylinder for the viscosity $\mu$ and pressure $p_t$ is calculated as
\begin{align} \label{eq:drag}
\mathbf{F}_{\mu} &= \oiint \mu (\nabla \mathbf{u} + \nabla \mathbf{u}^\intercal) \cdot \mathbf{n} \,dS, \\
\mathbf{F}_p &= \oiint -p \mathbf{n} \,dS, \\
C_{d, \mu} &= \frac{2 \cdot \mathbf{F}_{\mu} \cdot \mathbf{u}_{\infty}}{\varrho \cdot \| \mathbf{u}_{\infty} \|^3 \cdot D}, \\
C_{d, p} &= \frac{2 \cdot \mathbf{F}_p \cdot \mathbf{u}_\infty}{\varrho \cdot \| \mathbf{u}_{\infty} \|^3 \cdot D}, \\
C_{d}&= C_{d, \mu} + C_{d, p},
\end{align}
where $\mathbf{u}_\infty = (1, 0)^\intercal$ is the free-stream velocity and $\mathbf{n}$ is the outward normal of the cylinder perimeter.

The state of the LED at every time-step is composed of four fields,  the two components of the velocity field $u_x$, and $u_y$, the scalar pressure $p$ at each grid-point, and the vorticity field $\omega$ computed a-posteriori from the velocity field, i.e. $\mathbf{s}_t = \{ u_x, u_y, p, \omega\} \in \mathbb{R}^{4 \times 512 \times 1024}$.
The simulation state $\mathbf{s}_t$ is saved at a coarse time resolution $\Delta t=0.2$ for a total of $1000$ coarse time-steps.
There are $512$ grid points along the length of the channel and $1024$ gird points along the width of the channel.
After discarding the initial transients, $250$ time-steps are used for training (equivalent to $T=50$ time units), the next $250$ for validation  (equivalent to $T=50$ time units), and the next $500$ for testing  (equivalent to $T=100$ time units).

\begin{figure}[tbhp]
\centering
\includegraphics[width=0.9\textwidth]{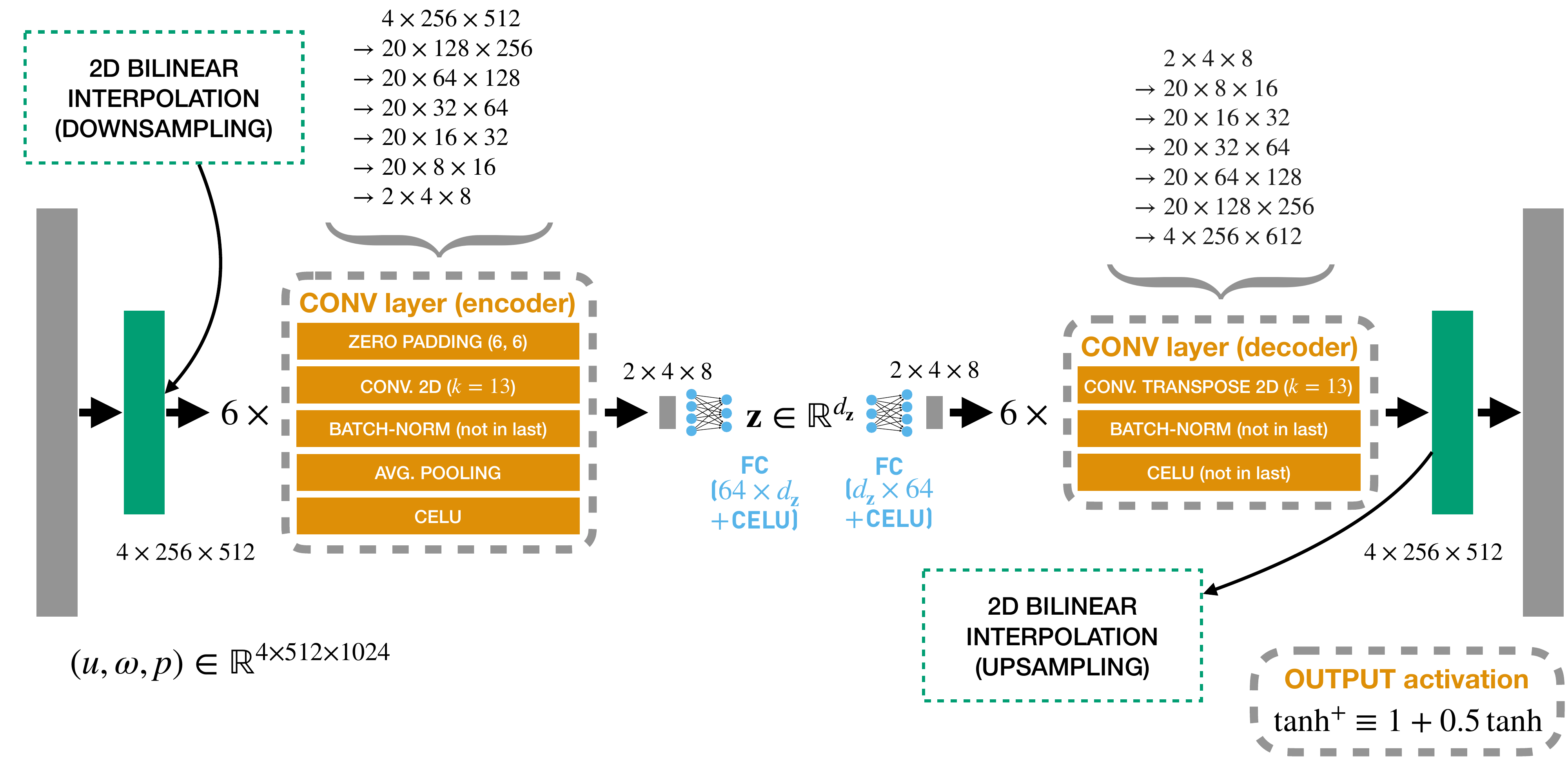}
\caption{
The architecture of the CNN employed in the flow behind a cylinder example.
}
\label{fig:cylRe_CNN}
\end{figure}

LED employs a Convolutional neural network (CNN) to identify a low dimensional latent space $\mathbf{z} \in \mathbb{R}^4$ in the $Re=100$ scenario, and $\mathbf{z} \in \mathbb{R}^{10}$ in the $Re=1000$ scenario.
The CNN architecture is depicted in~\ref{fig:cylRe_CNN}, and the layers are given in Table~\ref{app:tab:ns:cnn}.
We experimented with various activation functions, addition of batch-normalization layers, addition of transpose convolutional layers in the decoding part, different kernel sizes, and optimizers.
The data are scaled to $[0,1]$.
The output activation function of the CNN autoencoder is set to $0.5+0.5 \tanh(\cdot)$, whose image range matches the data range.

\begin{table}[tbhp]
\caption{CNN Autoencoder of LED for the flow behind a cylinder at $Re\in \{100, 1000\}$}
\label{app:tab:ns:cnn}
\centering
\resizebox{0.75\textwidth}{!}{
\begin{tabular}{ |c|c| } 
\hline
Layer & \text{ENCODER}  \\ \hline 
(0) & interpolationLayer() \\ \hline 
(1) & ZeroPad2d(padding=(6, 6, 6, 6), value=0.0) \\ \hline 
(2) & Conv2d(4, 20,  kernel\_size=(13, 13), stride=(1, 1)) \\ \hline 
(3) & BatchNorm2d(20, eps=1e-05, momentum=0.1, affine=0, track\_running\_stats=True) \\ \hline 
(4) & AvgPool2d( kernel\_size=2, stride=2, padding=0) \\ \hline 
(5) & CELU(alpha=1.0) \\ \hline 
(6) & ZeroPad2d(padding=(6, 6, 6, 6), value=0.0) \\ \hline 
(7) & Conv2d(20, 20,  kernel\_size=(13, 13), stride=(1, 1)) \\ \hline 
(8) & BatchNorm2d(20, eps=1e-05, momentum=0.1, affine=0, track\_running\_stats=True) \\ \hline 
(9) & AvgPool2d( kernel\_size=2, stride=2, padding=0) \\ \hline 
(10) & CELU(alpha=1.0) \\ \hline 
(11) & ZeroPad2d(padding=(6, 6, 6, 6), value=0.0) \\ \hline 
(12) & Conv2d(20, 20,  kernel\_size=(13, 13), stride=(1, 1)) \\ \hline 
(13) & BatchNorm2d(20, eps=1e-05, momentum=0.1, affine=0, track\_running\_stats=True) \\ \hline 
(14) & AvgPool2d( kernel\_size=2, stride=2, padding=0) \\ \hline 
(15) & CELU(alpha=1.0) \\ \hline 
(16) & ZeroPad2d(padding=(6, 6, 6, 6), value=0.0) \\ \hline 
(17) & Conv2d(20, 20,  kernel\_size=(13, 13), stride=(1, 1)) \\ \hline 
(18) & BatchNorm2d(20, eps=1e-05, momentum=0.1, affine=0, track\_running\_stats=True) \\ \hline 
(19) & AvgPool2d( kernel\_size=2, stride=2, padding=0) \\ \hline 
(20) & CELU(alpha=1.0) \\ \hline 
(21) & ZeroPad2d(padding=(6, 6, 6, 6), value=0.0) \\ \hline 
(22) & Conv2d(20, 20,  kernel\_size=(13, 13), stride=(1, 1)) \\ \hline 
(23) & BatchNorm2d(20, eps=1e-05, momentum=0.1, affine=0, track\_running\_stats=True) \\ \hline 
(24) & AvgPool2d( kernel\_size=2, stride=2, padding=0) \\ \hline 
(25) & CELU(alpha=1.0) \\ \hline 
(26) & ZeroPad2d(padding=(6, 6, 6, 6), value=0.0) \\ \hline 
(27) & Conv2d(20, 2,  kernel\_size=(13, 13), stride=(1, 1)) \\ \hline 
(28) & AvgPool2d( kernel\_size=2, stride=2, padding=0) \\ \hline 
(29) & CELU(alpha=1.0) \\ \hline 
(30) & Flatten(start\_dim=-3, end\_dim=-1) \\ \hline 
(31) & Linear(in\_features=64, out\_features=$d_{\mathbf{z}}$, bias=True) \\ \hline 
(32) & CELU(alpha=1.0) \\ \hline 
& $\mathbf{z} \in \mathbb{R}^{d_{\mathbf{z}}}$\\ \hline  \hline
Layer & \text{DECODER}  \\ \hline 
(0) & Linear(in\_features=$d_{\mathbf{z}}$, out\_features=64, bias=True)  \\ \hline 
(1) & CELU(alpha=1.0)  \\ \hline 
(2) & ViewModule()  \\ \hline 
(3) & ConvTranspose2d(2, 20,  kernel\_size=(13, 13), stride=(2, 2), padding=(6, 6), output\_padding=(1, 1))  \\ \hline 
(4) & BatchNorm2d(20, eps=1e-05, momentum=0.1, affine=0, track\_running\_stats=False)  \\ \hline 
(5) & CELU(alpha=1.0)  \\ \hline 
(6) & ConvTranspose2d(20, 20,  kernel\_size=(13, 13), stride=(2, 2), padding=(6, 6), output\_padding=(1, 1))  \\ \hline 
(7) & BatchNorm2d(20, eps=1e-05, momentum=0.1, affine=0, track\_running\_stats=False)  \\ \hline 
(8) & CELU(alpha=1.0)  \\ \hline 
(9) & ConvTranspose2d(20, 20,  kernel\_size=(13, 13), stride=(2, 2), padding=(6, 6), output\_padding=(1, 1))  \\ \hline 
(10) & BatchNorm2d(20, eps=1e-05, momentum=0.1, affine=0, track\_running\_stats=False) \\ \hline 
(11) & CELU(alpha=1.0)  \\ \hline 
(12) & ConvTranspose2d(20, 20,  kernel\_size=(13, 13), stride=(2, 2), padding=(6, 6), output\_padding=(1, 1))  \\ \hline 
(13) & BatchNorm2d(20, eps=1e-05, momentum=0.1, affine=0, track\_running\_stats=False) \\ \hline 
(14) & CELU(alpha=1.0)  \\ \hline 
(15) & ConvTranspose2d(20, 20,  kernel\_size=(13, 13), stride=(2, 2), padding=(6, 6), output\_padding=(1, 1))  \\ \hline 
(16) & BatchNorm2d(20, eps=1e-05, momentum=0.1, affine=0, track\_running\_stats=False) \\ \hline 
(17) & CELU(alpha=1.0)  \\ \hline 
(18) & ConvTranspose2d(20, 4,  kernel\_size=(13, 13), stride=(2, 2), padding=(6, 6), output\_padding=(1, 1))  \\ \hline 
(19) & interpolationLayer()  \\ \hline 
(20) & 1 + 0.5 Tanh() \\
\hline  \hline 
Latent dimension $d_{\mathbf{z}}$  & $\{1,2, 3, 4, 5, 6, 7, 8, 9, 10, 11, 12, 16\}$ \\ \hline
\end{tabular}
}
\\[0.4cm]
\begin{tabular}{ |c|c|c| } 
\multicolumn{3}{c}{Training times [minutes]} \\ \hline
\text{Min} & \text{Mean} & \text{Max} \\  \hline \hline
1080 & 1081 & 1083 \\
\hline
\end{tabular}
\end{table}


The hyper-parameter tuning and training times for the LSTM-RNN of LED are given in~\Cref{app:tab:ns:cnnrnns}.
The hyper-parameters and training times for the RC are given in~\Cref{app:tab:hyprc_ns}.
The hyper-parameters and training times for SINDy are given in~\Cref{app:tab:hypsindy_ns}.

\begin{table}[tbhp]
\caption{LED (LSTM-RNN) hyper-parameters and training times for the flow behind a cylinder example}
\label{app:tab:ns:cnnrnns}
\centering
\begin{tabular}{ |c|c|c| } 
\hline
\text{Hyperparameter} & 
\text{Values} \\  \hline \hline
Optimizer& Adabelief \\
Batch size& $32$ \\
Initial learning rate & $0.001 $ \\
Max Epochs& $1000$ \\
BPTT sequence length $L$ & $\{ {\color{red}10}, 25 \}$ \\
Warm-up steps & $10$ \\
Prediction horizon & $ 1000 $ \\
RNN Cell & LSTM \\
Number of RNN layers & $1$ \\
Size of RNN layers & $\{ {\color{red}32}, 64 \}$ \\
Scaling & $[0,1]$ \\
\hline 
\end{tabular}
\\[0.4cm]
\begin{tabular}{ |c|c|c| } 
\multicolumn{3}{c}{Training times [minutes]} \\ \hline
\text{Min} & \text{Mean} & \text{Max} \\  \hline \hline
722 & 723 & 724 \\
\hline
\end{tabular}
\end{table}

The hyper-parameters and training times for the RC are given in~\Cref{app:tab:hyprc_ns}.
The hyper-parameters and training times for SINDy are given in~\Cref{app:tab:hypsindy_ns}.

\begin{table}[tbhp]
\caption{Reservoir Computer hyper-parameters and training times (in CNN-RC) for flow behind a cylinder example}
\label{app:tab:hyprc_ns}
\centering
\begin{tabular}{ |c|c|c| } 
\hline
\text{Hyper-parameter tuning} & 
\text{Values for $Re=100$} & 
\text{Values for $Re=1000$} \\  \hline \hline
Solver & Pseudoinverse& Pseudoinverse \\
Size & $200$ & $200$ \\
Degree & $10$ & $10$ \\
Radius & $0.99$ & $0.99$ \\
Input scaling $\sigma$ & $\{0.5,  1, {\color{red} 2} \}$ & $\{ {\color{red} 0.5} ,  1, 2\}$  \\
Dynamics length & $100$ & $100$ \\
Regularization $\eta$ & $\{  0.0, 0.001, {\color{red} 0.0001 } , 0.00001\}$ & $\{  0.0,{\color{red} 0.001}, 0.0001,0.00001 \}$ \\
Noise level per mill & $\{  10 \}$ & $\{  10 \}$ \\
\hline
\end{tabular}
\\[0.4cm]
\begin{tabular}{ |c|c|c| } 
\multicolumn{3}{c}{Training times [minutes]} \\ \hline
\text{Min} & \text{Mean} & \text{Max} \\  \hline \hline
1.2 & 1.4 & 1.92 \\
\hline
\end{tabular}
\end{table}


\begin{table}[tbhp]
\caption{SINDy hyper-parameters and training times (in CNN-SINDy) for flow behind a cylinder example}
\label{app:tab:hypsindy_ns}
\centering
\begin{tabular}{ |c|c|c| } 
\hline
\text{Hyper-parameter tuning} & 
\text{Values for $Re=100$}& 
\text{Values for $Re=1000$} \\  \hline \hline
Library & Polynomials & Polynomials \\
Degree & $\{ {\color{red} 1},2,  3\}$ & $\{ {\color{red} 1},2,  3\}$ \\
Threshold & $\{ {\color{red} 0.001}, 0.0001, 0.00001 \}$ & $\{ 0.001, 0.0001, {\color{red} 0.00001} \}$ \\
\hline
\end{tabular}
\\[0.4cm]
\begin{tabular}{ |c|c|c| } 
\multicolumn{3}{c}{Training times [minutes]} \\ \hline
\text{Min} & \text{Mean} & \text{Max} \\  \hline \hline
1.14 & 1.55 & 2.05 \\
\hline
\end{tabular}
\end{table}



\clearpage
\newpage

\subsection{Alanine Dipeptide Dynamics}
\label{sec:alanine}

\begin{figure*}[tbhp]
\includegraphics[width=0.9\textwidth]{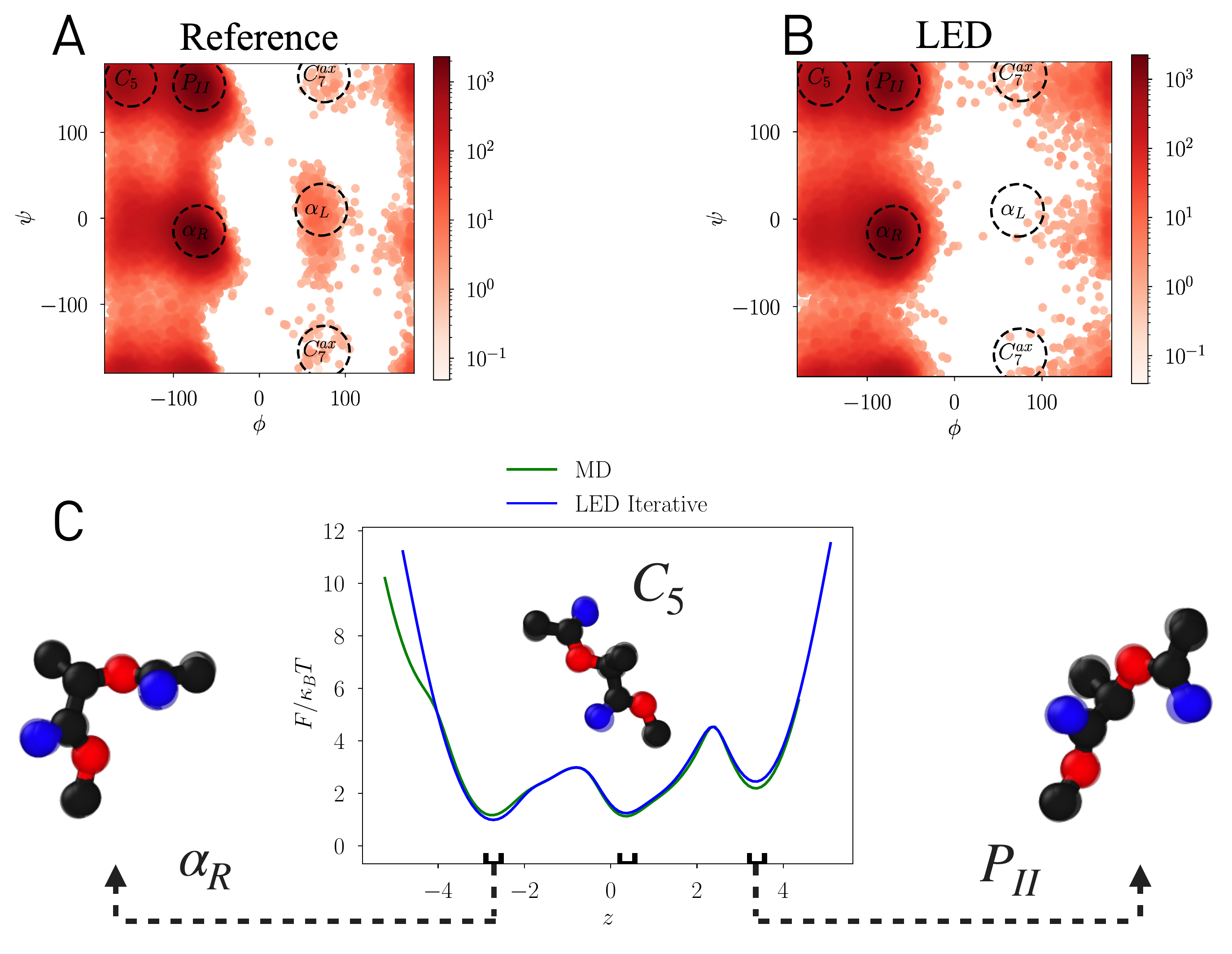}
\caption{
\textbf{A)}
Ramachadran plot of the alanine dipeptide data, i.e. state density on the space spanned by two backbone dihedral angles $(\phi, \psi)$.
\textbf{B)}
Ramachadran plot of the state evolution data predicted by LED with $T_{\mu}=0$ and $d_{\bm{z}}=1$.
LED captures the three mostly visited meta-stable states $\{C_5, P_{II}, \alpha_R\}$.
\textbf{C)} 
Projection of the state evolution data to the free energy on the one dimensional latent space unraveled by LED, i.e. $F /\kappa_B T=-\log \, p(z_t)$.
Low energy (high probability) regions on the latent space are mapped to known metastable state configurations of alanine dipeptide.
}
\label{fig:alanine:alanine}
\end{figure*}

The efficiency of LED in capturing complex molecular dynamics is demonstrated in the dynamics of a molecule of alanine dipetide in water, a benchmark for enhanced sampling methods.
The molecule is simulated with molecular dynamics with a time-step $\delta t = 1 \text{fs}$, to generate data of total length $38.4\text{ns}$ for training, $38.4\text{ns}$ for validation, and $100\text{ns}$ for testing.
The time-step of LED is set to $\Delta t=0.1\text{ps}$.
In this case, LED utilizes an MD decoder, and an MD-LSTM in the latent space, to model the stochastic, non-Markovian latent dynamics.
The latent space dimension is set to $d_{\bm{z}}=1$.
As shown in Figure~\ref{fig:alanine:alanine}, LED identifies a meaningful one dimensional latent space, and reproduces statistics of the system.
In our recent work~\cite{vlachas2021accelerated}, we demonstrate that LED also captures the time-scales between the meta stable states and samples realistic protein configurations while being three orders of magnitude faster than the molecular dynamics solver.

\end{document}